\documentclass[aps,preprint,nofootinbib,preprintnumbers,eqsecnum,superscriptaddress]{revtex4}
\pdfoutput=1
\usepackage{wrapfig}

\usepackage{color}

\usepackage[
      colorlinks=true,
      linkcolor=blue,
      urlcolor=blue,
      filecolor=black,
      citecolor=red,
      pdfstartview=FitV,
      pdftitle={},
        pdfauthor={Alejandra Castro, Stephane Detournay, Nabil Iqbal, Eric Perlmutter},
        pdfsubject={},
        pdfkeywords={},
        pdfpagemode=None,
        bookmarksopen=true
      ]{hyperref}

\usepackage[font=footnotesize,labelfont=bf,justification=centerlast,width=.94\textwidth]{caption}

\usepackage[normalem]{ulem}
\usepackage{amsmath}
\usepackage{enumerate}
\usepackage{amsfonts}
\usepackage{yfonts}

\usepackage{subfigure}
\usepackage{psfrag}

\usepackage{epsfig}
\usepackage[latin1]{inputenc}
\usepackage{float}
\usepackage{graphicx}
\usepackage{cancel}
\usepackage{mathrsfs}
\usepackage{amssymb}
\usepackage{amsfonts}
\usepackage{amsmath}
\usepackage{slashed}

\usepackage{graphicx}
\usepackage{bm}

\def\({\left(}
\def\){\right)}
\def\[{\left[}
\def\]{\right]}
\def\<{\langle}
\def\>{\rangle}

\newcommand{\bmat}{\begin{bmatrix}}
\newcommand{\emat}{\end{bmatrix}}

\def\Tr{\mathop{\rm Tr}}

\newcommand\half{{\ensuremath{\frac{1}{2}}}}
\newcommand\p{\ensuremath{\partial}}

\newcommand\field[1]{{\ensuremath{\mathbb{{#1}}}}}

\newcommand{\RR}{\field{R}}

\newcommand{\be}{\begin{equation}}
\newcommand{\ee}{\end{equation}}
\newcommand{\bea}{\begin{eqnarray}}
\newcommand{\eea}{\end{eqnarray}}
\newcommand{\bwt}{\begin{widetext}}
\newcommand{\ewt}{\end{widetext}}

\newcommand{\bi}{\begin{itemize}}
\newcommand{\ei}{\end{itemize}}
\newcommand{\ben}{\begin{enumerate}}
\newcommand{\een}{\end{enumerate}}
\newcommand{\bca}{\begin{cases}}
\newcommand{\eca}{\end{cases}}
\newcommand{\bln}{\begin{align}}
\newcommand{\eln}{\end{align}}
\newcommand{\bst}{\begin{split}}
\newcommand{\est}{\end{split}}

\newcommand\al{{\alpha}}
\newcommand\ep{\epsilon}
\newcommand\sig{\sigma}

\newcommand\lam{\lambda}
\newcommand\Lam{\Lambda}

\newcommand\Ga{{\ensuremath{{\Gamma}}}}

\def\th{{\theta}}

\newcommand\ha{{\half}}

\def\le{\left}
\def\ri{\right}

\newcommand\sG{{\ensuremath{{\mathcal G}}}}

\newcommand\sM{{\ensuremath{{\mathcal M}}}}

\newcommand\sO{{\ensuremath{{\mathcal O}}}}
\newcommand\sR{{\ensuremath{{\mathcal R}}}}

\newcommand{\ka}{{\kappa}}

\newcommand{\slt}{SL(2,\RR)}

\newcommand\zb{\bar{z}}

\newcommand\fm{\mathfrak{m}}
\newcommand\fs{\mathfrak{s}}
\newcommand\tn{\tilde{n}}
\newcommand\tq{\tilde{q}}

\def\rar{\rightarrow}
\def\C{\mathcal{C}}

\def\zb{\overline{z}}
\def\Tb{\overline{T}}
\def\bt{\beta}

\def\cP{{\cal P}}
\def\cR{{\cal R}}
\def\wb{\overline{w}}
\def\lan{\langle}
\def\ran{\rangle}
\def\ub{\overline{u}}
\def\vb{\overline{v}}

\begin{document}

\title {Holographic entanglement entropy and gravitational anomalies}


\author{Alejandra Castro}
\affiliation{Institute for Theoretical Physics, University of Amsterdam,
Science Park 904, Postbus 94485, 1090 GL Amsterdam, The Netherlands}

 \author{Stephane Detournay}
\affiliation{Physique Th\'eorique et Math\'ematique,
Universit\'e Libre de Bruxelles and International Solvay Institutes, Campus Plaine C.P.~231, B-1050 Bruxelles, Belgium}

\author{Nabil Iqbal}
\affiliation{Kavli Institute for Theoretical Physics, University of California, Santa Barbara CA 93106  }

\author{Eric Perlmutter}
\affiliation{DAMTP, Centre for Mathematical Sciences,  University of Cambridge, CB3 0WA, UK}

\begin{abstract}
\vskip 0.3in
We study entanglement entropy in two-dimensional conformal field theories with a gravitational anomaly. In theories with gravity duals, this anomaly is holographically represented by a gravitational Chern-Simons term in the bulk action. We show that the anomaly broadens the Ryu-Takayanagi minimal worldline into a ribbon, and that the anomalous contribution to the CFT entanglement entropy is given by the twist in this ribbon. The entanglement functional may also be interpreted as the worldline action for a spinning particle -- that is, an anyon -- in three-dimensional curved spacetime. We demonstrate that the minimization of this action results in the Mathisson-Papapetrou-Dixon equations of motion for a spinning particle in three dimensions. We work out several simple examples and demonstrate agreement with CFT calculations. 

\end{abstract}

\vfill

\today

\maketitle

\tableofcontents

\section{Introduction}
Holography relates gravitational physics to the dynamics of quantum field theories with a large number of degrees of freedom. The precise mechanism governing the emergence of an effective geometry from field-theoretical degrees of freedom remains somewhat unclear. However, recent work has shown that the concept of quantum entanglement is likely to play a key role in an eventual understanding of this emergence. Consider the {\it entanglement entropy} of a spatial region $X$ in the quantum field theory, defined as the von Neumann entropy of the reduced density matrix $\rho_X$ characterizing the spatial region. This is a very complicated and nonlocal observable. However, in field theories with semiclassical gravity duals governed by Einstein gravity, there exists a simple and elegant formula for the entanglement entropy, due to Ryu and Takayanagi \cite{Ryu:2006bv}:
\be
S_{\rm EE} = \frac{A_{\rm min}}{4 G_N},
\ee
where $A_{\rm min}$ denotes the area of the bulk minimal surface anchored on the boundary of the spatial region $X$. This equation relates two very primitive ideas on either side of the duality -- geometry and entanglement -- and thus may provide insight into the reorganization of degrees of freedom implied by holography. 

In this paper we study holographic entanglement entropy in a different class of theories, namely, two-dimensional conformal field theories with {\it gravitational anomalies} \cite{AlvarezGaume:1983ig}. These theories suffer from a breakdown of stress-energy conservation at the quantum level: this may be understood as a sensitivity of the theory to the coordinate system used to describe the manifold on which it is placed. The anomaly is present in any theory with unequal left and right central charges appearing in the local conformal algebras. If such a theory admits a gravity dual, the holographic three-dimensional bulk action will contain a gravitational Chern-Simons term. 

We have two main motivations in mind for studying entanglement entropy in such theories: from a field-theoretical point of view, we would like to better understand the interplay between anomalies and entanglement entropy. Indeed the celebrated formula $S_{\rm EE} = \frac{c}{3}\log \frac{R}{\ep}$ for the entanglement entropy of an interval in the vacuum of a two-dimensional conformal field theory is an example of a precise relation between these two concepts, but is controlled by the {\it conformal} anomaly instead. Another motivation is holographic: as we will see, in theories with anomalies, entanglement entropy probes other aspects of the dual bulk geometry. 

We find that the sensitivity of the field theory to its coordinate system manifests itself holographically in an elegantly geometric way: it gives physical meaning to a normal frame attached to the Ryu-Takayanagi worldine, essentially broadening the minimal worldline into a {\it ribbon}. The anomalous contribution to the entanglement entropy is now given by the twist in this ribbon:
\be
S_{\rm anom} = \frac{1}{4 G_3 \mu} \int_{C} ds \le( \tilde{n} \cdot \nabla n \ri), \label{Sanom1}
\ee
where $C$ is a curve in the three dimensional bulk and the vectors $n$ and $\tilde{n}$ define a normal frame to this curve. $\mu^{-1}$ appears in the coefficient of the gravitational Chern-Simons term, and measures the anomaly coefficient in the dual two-dimensional theory. 

Interestingly, just as a the length of a bulk spatial geodesic may be understood as the action of a massive point particle propagating in the bulk, this new term may be understood in terms of the action of a {\it spinning} particle, i.e. an {\it anyon} in AdS. Boundary intuition for this result comes from the fact that in two-dimensional conformal field theories with a gravitational anomaly, the twist fields used to compute entanglement entropy acquire nonzero spin proportional to the anomaly coefficient. We discuss the bulk interpretation of this action in detail, demonstrating in particular that its variation results in the usual Mathisson-Papapetrou-Dixon equations of motion for a spinning particle in general relativity \cite{Mathisson:1937zz,Papapetrou:1951pa,Dixon:1970zza}. Thus our results may also be viewed as the construction of the worldline action of an anyon in curved space. We anticipate further applications of this formalism. 

A brief outline of the paper follows. In Section \ref{sec:cft} we use two-dimensional conformal field theory techniques to study R\'enyi and entanglement entropy in field theories with gravitational anomalies; no use is made of holographic duality in that section. In Section \ref{sec:holEE} we derive the bulk result \eqref{Sanom1} by a careful evaluation of the action of a cone in topologically massive gravity, following techniques developed in \cite{Lewkowycz:2013nqa}. We also discuss its interpretation in terms of the action of a spinning particle and the relation to the Mathisson-Papapetrou-Dixon equations. In Section \ref{sec:results} we apply \eqref{Sanom1} to various simple spacetimes of interest, holographically deriving results that agree with the field theory calculations from Section \ref{sec:cft}. We conclude with a brief discussion and summary of future directions in Section \ref{sec:conc}. Many important details of the derivations have been relegated to various appendices. 

Appendix \ref{app:CS} presents a very different reformulation of the results of this paper, using the approach studied in \cite{Ammon:2013hba} to instead compute entanglement entropy from bulk Wilson lines in the Chern-Simons formulation of topologically massive gravity. 

Previous discussion of entanglement entropy in topologically massive gravity includes \cite{Sun:2008uf,Alishahiha:2013zta}: we believe that these works are incomplete in that they do not deal sufficiently carefully with the subtle lack of diffeomorphism invariance expected in these theories.

\section{Quantum entanglement in 2d CFTs with gravitational anomalies}\label{sec:cft}

\subsection{Gravitational anomalies}

We begin by recalling some basics about gravitational anomalies in 2d conformal field theories (CFTs). More complete accounts of the subject, including generalization to higher dimensions, can be found in e.g. \cite{AlvarezGaume:1983ig, Ginsparg:1985qn, Jensen:2013kka}.

A 2d CFT has Virasoro symmetries on the left and right that are generated by holomorphic and anti-holomorphic components of the stress tensor, $T(z)\equiv 2\pi T_{zz}$ and $\overline{T}(\zb)\equiv 2\pi T_{\zb\zb}$, respectively. 
The algebras take the familiar forms
\bea
T(z)T(0) &\sim& {c_L/2\over z^4} + {T(0)\over z^2}+{\p T(0)\over z} + \ldots~,\cr  && \cr
\Tb(\zb)\Tb(0) &\sim& {c_R/2\over \zb^4} + {\Tb(0)\over \zb^2}+{\overline{\p}\Tb(0)\over \zb} + \ldots~.
\eea
Associated to each algebra is an independent central charge. $E_L$ and $E_R$ will denote the left and right Virasoro zero modes, respectively. 

One can couple any QFT to a background metric $g_{ij}$ which acts as a source for the stress tensor $T_{ij}$, where $(i,j)$ run over the spacetime coordinates $(z,\zb)$. In terms of the QFT generating functional $W=-\log Z$, the stress tensor is defined as
\be
T_{ij} = {2\over \sqrt{g}}{\delta W\over \delta g^{ij}}~.
\ee
If $c_L\neq c_R$ in a 2d CFT, one cannot consistently promote the background metric to a dynamical field: that is, the CFT suffers from a gravitational anomaly. Examples include theories with chiral matter and holomorphic CFTs \cite{monster:1984,Witten:2007kt}. 

The anomaly so defined can be presented in two ways. On the one hand, it is manifest as a diffeomorphism anomaly, in which case the stress tensor is symmetric but not conserved. Under an infinitesimal diffeomorphism generated by $\xi_{\mu}$, the metric transforms as
\be
\delta g_{ij} = \nabla_{(i}\xi_{j)}~.
\ee
Non-invariance of the CFT generating functional implies
\be\label{stressnc}
\nabla^{i}T_{ij} =\left({c_L-c_R\over 96\pi}\right) \epsilon^{kl}\p_{k}\p_{m}\Gamma^{m}_{jl}~.
\ee
On the other hand, the anomaly can appear as a Lorentz anomaly -- that is, an anomaly under local frame rotations -- in which case the stress tensor is conserved but not symmetric. Working in the tangent frame, we have the vielbein $e^a_{~i}$, spin connection $\omega^a_{~b,i}$ and curvature 2-form $R_{ab}$, where $(a,b)$ are frame indices; see \cite{Ginsparg:1985qn} for a fuller description of the passage between metric and frame formulations. Under an infinitesimal frame rotation, the vielbein transforms as
\be
\delta e^{a}_{~i} = -\alpha^a_{~b}e^b_{~i}
\ee
with $\alpha_{ab}=-\alpha_{ba}$ the rotation parameter. Non-invariance of the CFT generating functional implies that this stress tensor, call it $\tilde{T}_{ij}$, obeys
\be
\tilde T_{ab}-\tilde T_{ba} = \left({c_L-c_R\over 48\pi}\right) {}^\star R_{ab}~,
\ee
where ${}^\star R_{ab}$ is the Hodge star of the curvature 2-form,
\be
R_{ab}=\half R_{ijkl}e^{~i}_ae^{~j}_bdx^{k}\wedge dx^{l}~.
\ee
The presentation of the anomaly is a matter of choice: shifting from one to the other can be done by adding a local counterterm to the CFT generating functional \cite{AlvarezGaume:1984dr,Bardeen:1984pm}.\footnote{This is succinctly captured in the language of the anomaly polynomial. For a gravitational anomaly in a 2d CFT, the polynomial is $\cP_4(R)=\Tr(R\wedge R)=d\omega_3$. 
The choice of diffeomorphism or Lorentz anomaly corresponds to the choice of whether to express the curvature as $R = d\Gamma + \Gamma \wedge \Gamma$ or $R = d\omega + \omega \wedge \omega$, respectively.}

The anomaly presented thus far is the so-called ``consistent'' form, obtained from a generating functional that satisfies the Wess-Zumino consistency conditions. There exists a ``covariant'' form of the anomaly, in which covariance of the stress tensor is restored via improvement terms, 
\be\label{imp}
\widehat{T}_{ij} = T_{ij} + Y_{ij}
\ee
for suitably chosen $Y_{ij}$ (the ``Bardeen-Zumino polynomials'') \cite{Bardeen:1984pm}. The covariant currents \eqref{imp} can be derived from a modified generating functional living in 2+1 dimensions. Framed as a diffeomorphism anomaly, for instance, the non-conservation equation in covariant form becomes
\be
\nabla^{i}\widehat{T}_{ij} =\left({c_L-c_R\over 96\pi}\right) \epsilon^{k}_{~j}\p_{k}R~,
\ee
where $R=R_{ij}g^{ij}$ is the Ricci scalar; note the absence of explicit factors of $\Gamma$. In a holographic context, the usual AdS/CFT dictionary identifying boundary data with sources in the CFT generating functional naturally leads to the consistent form \cite{Skenderis:2009nt}.

\subsection{R\'enyi and entanglement entropy in the presence of a gravitational anomaly}
Consider a QFT living on some Riemannian manifold $M$, in a state defined by some given density matrix. One can form a reduced density matrix, $\rho$, by tracing out degrees of freedom exterior to some spatial region at fixed Euclidean time. The R\'enyi entropy $S_n$ is defined as
\be\label{renyi}
S_{n}={1\over 1-n}\log {\rm Tr} \rho^n~,
\ee
where $n$ is a positive integer. In a 2d QFT, the spatial region is the union of $N\in \mathbb{Z}$  disjoint intervals. Upon analytically continuing $n$ to positive real values, one can take the limit $n\rar 1$ to obtain the entanglement, or von Neumann, entropy,
\be\label{eq:ee1}
S_{\rm EE}= \lim_{n\to 1}S_n = -\Tr\rho\log\rho~.
\ee
This procedure involves ambiguities in principle; we will employ the naive analytic continuation, based on precedent in similar 2d CFT entanglement calculations. For further details on this construction, we direct the reader to \cite{Calabrese:2004eu, Calabrese:2009qy, Headrick:2010zt}.

Computing entanglement entropy via the replica trick amounts to computing the R\'enyi entropy for arbitrary $n$. The latter is equivalent to computing a QFT partition function $Z_n$ on a branched cover of $M$, with branch cuts along the entangling surface. The precise relation is
\be\label{rhoz}
{\rm Tr} \rho^n = {Z_n\over (Z_1)^n}~.
\ee
This technique is applicable in any state that can be defined using a functional integral. Alternatively, one can eschew this topological perspective, and instead compute ${\rm Tr} \rho^n$ as a correlation function on $M$ of twist operators inserted at the boundary of the entangling region. 

We focus on 2d CFTs henceforth, calling a generic CFT $\C$. The branched covers of $M$ are Riemann surfaces of $N$- and $n$-dependent genus; when $M=\mathbb{C}$, for example, $g=(N-1)(n-1)$. Rather than computing partition functions on arbitrary genus Riemann surfaces directly, we will employ the twist field perspective \cite{Calabrese:2004eu}. For $N$ spatial intervals bounded by endpoints $\lbrace u_i,v_i\rbrace$, one inserts $N$ twist operators $\Phi_+$ at the endpoints $u_i$ and $N$ twist operators $\Phi_-$ at the endpoints $v_i$, and computes their $2N$-point correlator on $M$:
\be\label{twcor}
\Tr\rho^n = \left\langle \prod_{i=1}^N\Phi_+(u_i)\Phi_-(v_i)\right\rangle_{\C^n/\mathbb{Z}_n}~.
\ee
As the subscript denotes, the correlator is evaluated in the orbifold theory $\C^n/\mathbb{Z}_n$. $\Phi_+$ and $\Phi_-$ act oppositely with respect to the cyclic permutation symmetry $\mathbb{Z}_n$, and have a nontrivial OPE. The $n$-dependence also enters \eqref{twcor} through the scaling dimensions of the twist operators, to be derived momentarily. 

For a CFT with $c_L\neq c_R$, these techniques still apply. The presence of a gravitational anomaly does not spoil the use of functional integrals; one need only be careful about their transformation properties. This brings us to the central novelty: in a CFT with $c_L\neq c_R$, the twist fields $\Phi_{\pm}$ have nonzero spin controlled by the anomaly coefficient: $s\propto c_L-c_R$. 

To derive this fact, following the original method of \cite{Calabrese:2004eu}, it is sufficient to consider a single interval in the ground state of a CFT $\C$, with endpoints $(u,v)$. Before replication, $\C$ lives on the complex plane, $M=\mathbb{C}$. The replica surface -- call it $\cR_{n,1}$ -- has genus zero for all $n$, and can be conformally mapped back to the complex plane. If $w$ is a complex coordinate on $\cR_{n,1}$, then
\be\label{unif}
z= \left({w-u\over w-v}\right)^{1/n}~,
\ee
maps $\cR_{n,1}\mapsto \mathbb{C}$, coordinatized by $z$. To compute the scaling weights of the twist operators, consider the one-point functions of the CFT stress tensor components, $\lan T(w)\ran_{\cR_{n,1}}$ and $\lan \Tb(\wb)\ran_{\cR_{n,1}}$. On one hand, one can compute the conformal transformation of the stress tensor under the map \eqref{unif}. 
On the other, these expectation values can be traded for correlation functions with twist operators in the orbifold theory, by definition of the twist operators themselves. Recalling that $\lan T(z) \ran_{\mathbb{C}} = \lan \Tb(\zb) \ran_{\mathbb{C}} = 0$, one finds
\bea\label{tcor}
{\lan T(w)\Phi_+(u,\ub)\Phi_-(v,\vb) \ran_{\C^n/\mathbb{Z}_n}\over \lan \Phi_+(u,\ub)\Phi_-(v,\vb) \ran_{\C^n/\mathbb{Z}_n}} &=& {c_L\over 24}\left(n-{1\over n}\right)\left(u-v\over (w-u)(w-v)\right)^2~,\cr
{\lan \Tb(\wb)\Phi_+(u,\ub)\Phi_-(v,\vb) \ran_{\C^n/\mathbb{Z}_n}\over \lan \Phi_+(u,\ub)\Phi_-(v,\vb) \ran_{\C^n/\mathbb{Z}_n}} &=& {c_R\over 24}\left(n-{1\over n}\right)\left(\ub-\vb\over (\wb-\ub)(\wb-\vb)\right)^2~.
\eea
These equations are nothing but the conformal Ward identities for the holomorphic and anti-holomorphic stress tensors, acting on Virasoro primary operators with scaling dimensions
\be
h_L = {c_L\over 24}\left(n-{1\over n}\right)~, \quad h_R={c_R\over 24}\left(n-{1\over n}\right)~.
\ee
Defining conformal dimension $\Delta=h_L+h_R$ and spin $s = h_L-h_R$, we find spinning, anyonic twist fields as advertised:
\be\label{twdims}
\Delta = {c_L+c_R\over 24}\left(n-{1\over n}\right)~, \quad s={c_L-c_R\over 24}\left(n-{1\over n}\right)~.
\ee
This derivation is a simple extension of the case $c_L=c_R$ treated in \cite{Calabrese:2004eu} and elsewhere; the key point is that the calculation holomorphically factorizes. 

By deriving twist operator quantum numbers from the CFT stress tensor, this method fully accounts for the dependence of R\'enyi and entanglement entropy on the gravitational anomaly. We are implicitly working with the consistent form of the anomaly, because $T(z)$ and $\Tb(\zb)$ are defined through variations of the original CFT generating functional. Notice that in computing entanglement entropy as the $n\rar 1$ limit of R\'enyi entropy, the twist operators are analytically continued from anyons to spinless bosons. However, the ratio $s/\Delta$ is independent of $n$.

\subsection{Applications}
With these results in hand we proceed to easily derive universal results for single interval R\'enyi and entanglement entropy in 2d CFTs with arbitrary $(c_L,c_R)$. From \eqref{twcor}, this merely involves computing the two-point function of $\Phi_+$ and $\Phi_-$. For a CFT in its ground state, on a circle, or at finite temperature and angular potential, the functional form of the correlator is fixed by conformal symmetry \cite{DiFrancesco:1997nk}. We will also comment on more general entanglement entropies in holographic CFTs with $c_L\neq c_R$.

We work with an interval $[0, R]$ in a CFT on a line, with $R\in \mathbb{R}$, unless otherwise noted. Thus, we wish to compute
\be\label{eesing}
S_n={1\over 1-n}\log\lan \Phi_+(R)\Phi_-(0)\ran_{\C^n/\mathbb{Z}_n}~.
\ee
We will reinstate the UV cutoff as needed for dimensional analysis.
\subsubsection{Ground state}
In this case, $M=\mathbb{C}$, so \eqref{twdims} and \eqref{eesing} yield
\be\label{gdrenyi}
S_n = \left(1+{1\over n}\right){c_L+c_R\over 12}\log\left({R\over\varepsilon}\right)~.
\ee
and the entanglement entropy is easily obtained,
\be\label{gdee}
S_{EE} = {c_L+c_R\over 6}\log\left({R\over\varepsilon}\right)~.
\ee
This was first derived in \cite{Holzhey:1994we}. The anomaly makes no contribution to \eqref{gdee}. For holographic CFTs, this state is dual to Poincar\'e AdS.

\subsubsection{Ground state, boosted interval}
Now consider a complex interval $[0,z]$, with $z \in \mathbb{C}$ somewhere off the real axis. We find
\be
S_{n}= \left(1+{1\over n}\right)\left({c_L\over 12} \log \Big({z\over \varepsilon}\Big)+{c_R\over 12} \log \Big({\overline{z}\over \varepsilon}\Big)\right)~.
\ee
In this case, we {\it do} see a contribution from the anomaly. If we define $z = R e^{i\theta}$,
we find an entanglement entropy
\be\label{eevac0}
S_{\rm EE}= {c_L+c_R\over 6} \log \left({R\over \varepsilon}\right)+ {c_L-c_R\over 6} i\theta~.
\ee
This setup is just a rotation of the previous one, but in a theory with a Lorentz anomaly, observables are sensitive to the choice of frame. Equivalently, in the language of the diffeomorphism anomaly, observables are sensitive to the choice of coordinates used to define a constant time slice.

As \eqref{eevac0} shows, in the absence of Euclidean time reflection symmetry ($z\leftrightarrow \zb$) a Euclidean CFT partition function need not be real.\footnote{In fact, the anomalous piece had to be purely imaginary on general grounds. Thinking of $\Tr\rho^n$ as a CFT partition function on a Riemann surface, as in \eqref{rhoz}, we note that only the imaginary part of $\log Z$ can ever receive an anomalous contribution. Thus, the frame rotation to a complex interval must lead to an imaginary piece of the Euclidean R\'enyi entropy.} This is familiar from the thermal partition function of a non-anomalous CFT at finite temperature and chemical potential for angular momentum. As in that case, the complexity of \eqref{eevac0} is consistent with a sensible, real Lorentzian interpretation: under the analytic continuation $z=x-t,\zb=x+t$, the angle $\theta$ maps to a boost parameter $\kappa$ as $\theta=i\kappa$. Plugging this into \eqref{eevac0} yields a real  contribution to the entanglement entropy due to an anomalous Lorentz boost:
\be\label{eevac}
S_{\rm EE}= {c_L+c_R\over 6} \log \left({R\over \varepsilon}\right)- {c_L-c_R\over 6} \kappa~.
\ee

\begin{figure}[h]
\includegraphics[width=0.7\textwidth]{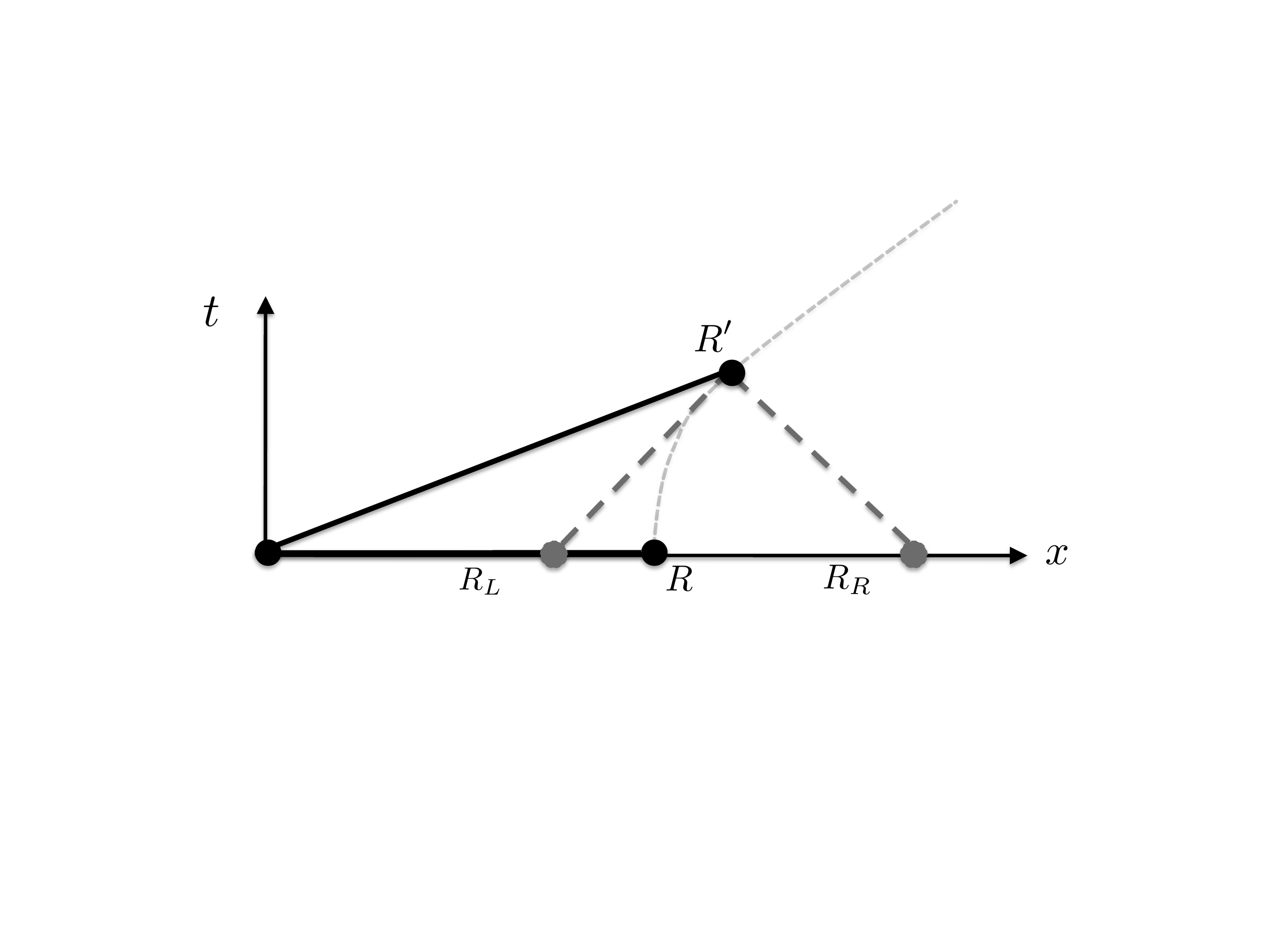}
\caption{Flow of entanglement through boosted interval.}
\label{fig:boost}
\end{figure}

There is a simple physical argument for this due to Wall \cite{Wall:2011kb}, as shown in Figure \ref{fig:boost}. 

To understand Figure \ref{fig:boost}, suppose that we have quantized our theory with respect to the time coordinate $t$; thus we understand how to compute entanglement entropy on the slice $t = 0$. Consider now the entanglement through the boosted interval (ending at $R'$). We would like to relate this to entanglement computed at $t = 0$. We see that at $t = 0$ all the left-movers that will eventually pass through the boosted interval are inside $R_L$, whereas all the right-movers are inside $R_R$. Assuming factorization of left and right-movers, the entanglement computed at $t = 0$ then should be
\be
S_{EE} = \frac{c_L}{6} \log \le(\frac{R_L}{\varepsilon}\ri) + \frac{c_R}{6} \log\le(\frac{R_R}{\varepsilon}\ri)
\ee
If the hyperbolic boost angle is $\ka$, then $R_L = R e^{-\ka}$ and $R_R = R e^{\ka}$: this immediately reproduces \eqref{eevac}.

\subsubsection{Finite temperature and angular potential}
Consider now a spatial interval $[0,R]$ in a general Lorentzian CFT on a line at finite temperature, $T=\bt^{-1}$, and chemical potential for angular momentum, $\Omega$. The CFT now lives on a cylinder with a compact thermal cycle, $M=S^1\times \mathbb{R}$. For a holographic CFT, this state is dual to the rotating, planar BTZ black hole. After Wick rotation, the Euclidean partition function is 
\be
Z=\Tr(e^{-\bt H -\bt\Omega_E J})
\ee
where 
\be
H=E_R+E_L-{c_L+c_R\over 24}~, \quad J=E_R-E_L+{c_L-c_R\over 24}~,
\ee
and we define $\Omega_E$ to be real via the standard analytic continuation $\Omega = i\Omega_E$. It is convenient to define left and right temperatures, $T_L\neq T_R$; the respective inverse temperatures $(\bt_L,\bt_R)$ are defined in terms of $(\bt,\Omega_E)$ as
\be\label{tempsss}
\bt_L = \bt(1+i\Omega_E)~, \quad \bt_R = \bt(1-i\Omega_E)~.
\ee
Notice that in the ground state on the cylinder, $E_L=E_R=0$, there is a nonzero ``Casimir momentum'' $J_0$ in addition to the usual ground state energy, $E_0$:
\be\label{gdst}
E_0 = -{c_L+c_R\over 24}~, \quad J_0={c_L-c_R\over 24}~.
\ee
This effect stems from the imbalanced chiral contributions to the Casimir energy.

To compute the R\'enyi entropy, we plug the universal form of cylinder correlation functions into \eqref{eesing}, which yields
\be\label{renyirot}
 S_{\rm n}=\left(1+{1\over n}\right)\left[{c_L\over 12}\log\left({\beta_L\over \pi \varepsilon}\sinh{\pi R\over \beta_L}\right)+{c_R\over 12}\log\left({\beta_R\over \pi \varepsilon}\sinh{\pi R\over \beta_R}\right)\right]~,
 \ee
and in the limit $n\to1$,
\be
\label{eerot}
 S_{\rm EE}={c_L\over 6}\log\left({\beta_L\over \pi \varepsilon}\sinh{\pi R\over \beta_L}\right)+{c_R\over 6}\log\left({\beta_R\over \pi \varepsilon}\sinh{\pi R\over \beta_R}\right)~.
 \ee
For $\bt_L\neq \bt_R$, there is an anomalous contribution that maps to a real entanglement entropy in Lorentzian signature. The formula neatly factorizes, and yields the correct Cardy entropy in the thermal limit $R\gg (\beta_L,\beta_R)$. In analogy to the case of the complex interval on the plane, we can think of $\Omega_E$ as a boost parameter, or as parameterizing passage to a rotating frame.

From the above results, we can also compute entanglement for a CFT at zero temperature on a cylinder of size $L$. We can read off the R\'enyi and entanglement entropies from \eqref{renyirot}-\eqref{eerot} by simply taking $\bt_L=\bt_R\mapsto -i L$. This case sees no anomalous contribution. For holographic CFTs, this state is dual to global AdS.

\subsubsection{Holographic CFTs at large central charge}
The results in this section universally apply to all CFTs with arbitrary $(c_L,c_R)$. For this (sub)subsection, we specialize to a class of ``holographic'' CFTs at large total central charge $c_L+c_R$: this should be understood as the class of CFTs which are likely to have semiclassical gravity duals, although no use will be made of holography in this section. A constructive definition of such CFTs was given in \cite{Headrick:2010zt, Hartman:2013mia}: aside from fundamental properties of unitarity, modular invariance and compactness, their essential property is that the OPE coefficients and density of states with $\Delta \lesssim c_L+c_R$ do not scale exponentially with large $c_L+c_R$. In anticipation of our study of holographic entanglement in TMG, this is an eminently relevant class of theories to consider. 

Recent work \cite{Headrick:2010zt, Hartman:2013mia, Faulkner:2013yia, Barrella:2013wja, Chen:2013kpa, Perlmutter:2013paa, Chen:2013dxa, Chen:2014kja} has shown that in such large central charge 2d CFTs with or without gravitational anomalies, R\'enyi and entanglement entropies can be derived for otherwise non-universal cases, involving higher genus manifolds $M$ or multiple intervals $N>1$. For example, one can think of \eqref{eerot} as the high temperature limit of entanglement entropy of a single interval on a torus, $M=T^2$, with complex modular parameter $\tau=i\bt_L/2\pi$. Torus two-point correlators are non-universal. For a CFT satisfying the above properties, however, the entanglement entropy of a single interval on the torus has been argued to be given by \eqref{eerot} for {\it all} temperatures above the Hawking-Page transition at leading order in large $c_L+c_R$ \cite{Barrella:2013wja, Perlmutter:2013paa}, provided that the size of the interval is smaller than half of the circle.\footnote{For larger interval sizes there may be a competing saddle, holographically manifest as the disconnected surface contributing to the Ryu-Takayanagi formula, i.e. the ``entanglement plateaux'' of \cite{Hubeny:2013gta}.} This provides a modest consistency check of \eqref{eerot}.

As a second example, consider the R\'enyi and entanglement entropy of {\it multiple} intervals in the ground state. As $2N$-point twist correlators with $N>1$, these are non-universal in general, but can be computed systematically in holographic CFTs in a large $c_L+c_R$ expansion \cite{Headrick:2010zt, Hartman:2013mia, Perlmutter:2013paa, Chen:2013kpa, Chen:2014kja}. This is true regardless of whether the CFT suffers from a gravitational anomaly. At leading order in large $c_L+c_R$, the ground state entanglement entropy is simply a sum over single interval contributions \eqref{gdee} subject to a certain minimization prescription \cite{Hartman:2013mia}:
\be
S_{EE} = {c_L+c_R\over 6}{\rm min}\sum_{(i,j)}\log\left({z_i-z_j\over\varepsilon}\right)~,
\ee
where $\lbrace z_i\rbrace$ label the endpoints, $i=1,\ldots ,2N$.

\section{Holographic entanglement entropy in topologically massive gravity} \label{sec:holEE}

We turn now to the holographic description of theories with gravitational anomalies. As mentioned earlier, the bulk action describing such a theory has a gravitational Chern-Simons term in its action. In trying to understand the anomalous contribution to entanglement, it is sufficient to work with {\it topologically massive gravity} (TMG), which couples the Chern-Simons term to ordinary Einstein gravity. The Ryu-Takayanagi prescription for computing entanglement entropy will require modification in this case. Roughly speaking, just as in ordinary 3d gravity the Ryu-Takayanagi worldline can be viewed as the trajectory traced out by a heavy bulk particle, we will show that in TMG the analog of this idea involves a massive {\it spinning} particle, whose on-shell action yields the yields the holographic entanglement entropy. This is nothing but the extension of the twist field spin into the bulk. Thus, our task is to construct an action for such a particle and understand its dynamics. As we will see, our action makes satisfying contact with previous literature on spinning particles in general relativity. 

We begin this section by briefly recalling some salient features of TMG.

\subsection{Gravitational Chern-Simons term}

The original work on TMG dates back to \cite{Deser:1981wh,Deser:1982vy,Deser:1991qk}; for a more modern treatment see e.g. \cite{Deser:2002iw,Li:2008dq, Skenderis:2009nt}.  The action is the sum of the Einstein-Hilbert term and a gravitational Chern-Simons term:
\bea\label{eq:atmg}
S_{\rm TMG}&=&{1\over 16\pi G_3}\int d^3x\sqrt{-g}\left(R+{2\over\ell^2}\right) \cr &&+{1\over 32\pi G_3\mu} \int d^3x\sqrt{-g}\epsilon^{\lambda\mu\nu}\Gamma^\rho_{\lambda\sigma}\left(\partial_\mu \Gamma^\sigma_{\rho\nu}+{2\over 3}\Gamma^\sigma_{\mu\tau}\Gamma^\tau_{\nu\rho}\right)~,
\eea
where we have included a negative cosmological constant. $\mu$ is a real coupling with dimensions of mass. Even though the action has explicit dependence on Christoffel symbols, the  equations of motion are covariant: explicitly, they are
\bea\label{eq:eomtmg}
R_{\mu\nu}-{1\over 2}g_{\mu\nu} R -{1\over \ell^2} g_{\mu\nu}=-{1\over \mu}C_{\mu\nu}~,
\eea
where $C_{\mu\nu}$ is the Cotton tensor,
\be\label{eq:cotton}
C_{\mu\nu}=\epsilon_\mu^{~\alpha\beta}\nabla_\alpha \left(R_{\beta\nu}-{1\over 4}g_{\beta\nu}R\right)~.
\ee
The Cotton tensor is symmetric, transverse and traceless,
\be\label{eq:cotton2}
\epsilon^{\alpha\mu\nu}C_{\mu\nu}=\nabla^{\mu}C_{\mu\nu} = C_{~\mu}^{\mu}=0~.
\ee
Consequently, all solutions of TMG have constant scalar curvature, $R=-6/\ell^2$, and the equations of motion can be written as
\be\label{eq:eom2}
R_{\mu\nu}+{2\over \ell^2}g_{\mu\nu} = -{1\over \mu}C_{\mu\nu}~.
\ee
The Cotton tensor thus measures deviations from an Einstein metric. 

It is clear from \eqref{eq:cotton}-\eqref{eq:eom2} that TMG admits all solutions of pure three dimensional gravity, namely, Einstein metrics for which $C_{\mu\nu}=0$. The theory also admits a wide class of non-Einstein metrics which are not locally AdS$_3$. We briefly note that TMG can also be cast in a Chern-Simons formulation; in Appendix \ref{app:CS} we review that approach to the theory.

Many unusual properties of TMG are put into sharper focus in a holographic context. This theory has been extensively studied in the context of AdS$_3$/CFT$_2$ by  \cite{Solodukhin:2005ns,Kraus:2005zm,Hotta:2008yq,Skenderis:2009nt}, among many other authors.\footnote{Holography for TMG has been also the cradle of some controversy at $\mu=1$, the so-called ``chiral point,'' which we will not discuss here. See e.g. \cite{Li:2008yz,Grumiller:2008qz,Maloney:2009ck} and references within.} Specializing for the moment to locally AdS$_3$ solutions, the application of Brown-Henneaux \cite{Brown:1986nw} for TMG shows that the classical phase space of asymptotically AdS$_3$ backgrounds is organized in two copies of the Virasoro algebra with central charges
\be\label{eq:clcr}
c_L={3\ell\over 2G_3}\left(1-{1\over \mu\ell}\right) ~,\quad c_R= {3\ell\over 2G_3}\left(1+{1\over\mu\ell}\right)~.
\ee
As detailed in the previous section, the unequal central charges indicate the presence of a gravitational anomaly in the dual CFT. This is consistent with the transformation of the gravitational Chern-Simons term under bulk diffeomorphisms: in particular, the bulk action is invariant up to a boundary term that captures the nonzero divergence of the CFT stress tensor in \eqref{stressnc} \cite{Kraus:2005zm}. This is the usual elegant mechanism of holographic anomaly generation in AdS/CFT \cite{Witten:1998qj} as applied to a gravitational anomaly.

One can also perform a precise matching of thermodynamic observables of the bulk and boundary theories. The presence of the gravitational Chern-Simons term generically affects the observables in the bulk, even for locally AdS$_3$ solutions. One intriguing example is the introduction of nonzero angular momentum into global AdS$_3$: the ADM charges are
\be
\ell M_{\rm AdS}= -{\ell\over 8G_3}=-{c_L+c_R\over 24}~,\quad J_{\rm AdS}=-{1\over 8G_3\mu}= {c_L-c_R\over 24}~,
\ee
which match the CFT result \eqref{gdst} for ground state energy and Casimir momentum on the cylinder \cite{Kraus:2005zm}. At finite temperature, the bulk admits BTZ black holes, which are dual to thermal states in the CFT. The asymptotic density of states of the CFT is governed by the Cardy formula \cite{Cardy:1986ie},
\be\label{cardy}
S = 2\pi\sqrt{c_L E_L\over 6} + 2\pi \sqrt{c_RE_R\over 6}~.
\ee
Happily, this has been shown to equal the Wald entropy of BTZ black holes in TMG or any covariant higher derivative modification thereof \cite{Wald:1993nt, Saida:1999ec, Kraus:2005vz, Solodukhin:2005ns,Tachikawa:2006sz, Sahoo:2006vz}.

The novel solutions of TMG are those with $C_{\mu\nu}\neq 0$, which are not locally AdS$_3$. An interesting subset of such solutions, denoted ``warped AdS$_3$'', were constructed in \cite{Moussa:2003fc,Bouchareb:2007yx,Anninos:2008fx}, along with warped  black hole counterparts. Viewed holographically, the main feature of asymptotically warped AdS$_3$ geometries is that they do not obey Brown-Henneaux boundary conditions, so \eqref{eq:clcr} is not applicable; indeed, the nature and symmetries of a putative holographic dual CFT are not very well understood (see, e.g., \cite{Guica:2011ia, Song:2011sr,ElShowk:2011cm,Detournay:2012pc,Compere:2008cv} and references within). 

\vskip .1in

In the following sections, we will deal with the Euclidean theory for both technical and conceptual reasons. The gravitational Chern-Simons term is a parity odd term under time reversal, hence a Wick rotation $t\to t_E=it$ yields 
\bea\label{eq:atmg2}
S_{\rm Euc}&=&{1\over 16\pi G_3}\int d^3x\sqrt{g}\left(R+{2\over\ell^2}\right) \cr &&+{i\over 32\pi G_3\mu} \int d^3x\sqrt{g}\epsilon^{\lambda\mu\nu}\Gamma^\rho_{\lambda\sigma}\left(\partial_\mu \Gamma^\sigma_{\rho\nu}+{2\over 3}\Gamma^\sigma_{\mu\tau}\Gamma^\tau_{\nu\rho}\right)~,
\eea
and the equations of motion are
\be\label{eq:eome}
R_{\mu\nu}+{2\over \ell^2}g_{\mu\nu} = -{i\over \mu}C_{\mu\nu}~.
\ee
The appearance of the factor of $i$ in \eqref{eq:eome} is important: for a real metric, the left and right hand side of \eqref{eq:eome} must vanish independently, hence restricting the space of real solutions to the Einstein metrics of pure gravity. Our presentation of the holographic entanglement functional for TMG relies only on general properties of the theory, modulo some mild assumptions; nevertheless, it is not completely obvious how to apply our result to the full set of solutions of TMG, for which the Euclidean continuation is unclear. We will comment on these subtleties in the Discussion section.

\subsection{Holographic entanglement entropy: Coney Island} \label{sec:coneyisland}

We now want to understand how the computation of holographic entanglement entropy in TMG reflects the gravitational anomaly of a dual CFT. In principle, there exists a perfectly well-defined procedure for computing R\'enyi entropy for any $n$. In \eqref{eq:ee1}-\eqref{rhoz} we wrote the entanglement entropy as
\be\label{eq:ee2}
S_{\rm EE}=\lim_{n\to 1}{1\over 1-n} (\log Z_n - n\log Z_1)~.
\ee
Using the AdS/CFT dictionary, we interpret $Z_n$ as the gravitational partition function for a 3-manifold ${\cal M}_n$ asymptotic to a replica manifold ${\cal R}_{n}\equiv\p {\cal M}_n$ at conformal infinity \cite{Headrick:2010zt,Faulkner:2013yia}. In the semiclassical limit we can approximate this partition function by computing the action of $\sM_n$.

In \cite{Lewkowycz:2013nqa} an efficient algorithm was developed for computing such actions in states that admit a description involving a Euclidean partition function. We do not review details here but just state the result: in the $n \to 1$ limit, one can understand the action of $\sM_n$ by studying the action of a bulk geometry with a conical surplus of $2\pi(n-1)$ along a worldline $C$ extending into the bulk, i.e.
\be
S_{\rm EE} = -\p_n(S_{\rm cone})\big|_{n=1} \ . \label{EEdef}
\ee
The Einstein-Hilbert action evaluated on this conical surplus simply measures the proper length of $C$. Furthermore one finds that consistency of the bulk equations requires (in the case of Einstein gravity) that the curve $C$ must be a bulk geodesic, and one eventually finds
\be\label{rt}
S_{\rm EE}^{(\rm RT)} = {L_{\rm min}\over 4G_3} \ . 
\ee
This constitutes a proof of the celebrated Ryu-Takayanagi formula (RT) for the entanglement entropy \cite{Ryu:2006bv,Ryu:2006ef,Nishioka:2009un}, at least for the set of states that can be studied using a Euclidean partition function. 

We would now like to apply these techniques to TMG. The fact that the bulk action is not {\it quite} gauge-invariant will play an interesting role in our analysis: the existence of this anomaly will effectively introduce new data along the worldline, broadening the usual Ryu-Takayanagi worldline into a {\it ribbon}. As we will demonstrate in detail, this is the bulk representation of the anomalous contributions to CFT entanglement discussed in Section \ref{sec:cft}, a statement we will verify in a number of examples. 

Consider a regularized cone in three-dimensional Euclidean space with opening angle $2\pi(1 + \epsilon)$, so that $\ep = n-1$. The tip of this cone defines a one-dimensional worldline that we take to extend along a spacelike direction parametrized by $y$. We use flat coordinates $\sig^a$ for the two perpendicular directions, and the tip of the cone is then at $\sig^a = 0$. A suitable metric ansatz for this cone is then
\be
ds^2 = e^{\ep \phi(\sig)} \delta_{ab} d\sig^a d\sig^b + \le(g_{yy} + K_a \sig^a + \cdots \ri) dy^2 + e^{\ep \phi(\sig)} U_a(\sig,y) d\sig^a dy \label{conemetnice}
\ee
In this parametrization the extrinsic curvatures $K_a$ are Taylor coefficients in the expansion of $g_{yy}$ in the $\sig^a$ directions, and they measure the deviation of the trajectory from a geodesic. $\phi(\sigma)$ stores the information of the regularization of the cone: importantly, it can be picked to fall off exponentially outside a small core.

We now compute the action of this cone \eqref{eq:atmg2} to first order in $\ep$, using standard techniques \cite{Lewkowycz:2013nqa,Dong:2013qoa,Camps:2013zua}, although the generalization to TMG requires some care. We present details of the calculation, including explicit expressions for the regulator function $\phi(\sig)$, in Appendix \ref{app:coneac}. 

Let us recall the logic by first evaluating the Einstein part of the action \eqref{eq:atmg2}. The Ricci scalar receives contributions of the form $\nabla_a^2 \phi$. Evaluating the integrals we find
\be
S_{\rm cone, Einstein} = -\frac{\ep}{4 G_3}\int_{C} dy \sqrt{g_{yy}} + {\cal O}(\epsilon^2)~.\label{noncovEin}
\ee
Although this expression was evaluated in the coordinate system given by \eqref{conemetnice}, it is clear that it is simply measuring the proper distance along the cone-tip worldline. In particular, there is no obstruction to writing a fully covariant expression: parametrizing the cone tip by $X^{\mu}(s)$, we simply find 
\be
S_{\rm cone, Einstein} = -\frac{\ep}{4G_3} \int_{C} ds \sqrt{g_{\mu\nu}(X) \dot{X}^{\mu} \dot{X}^{\nu}} + {\cal O}(\epsilon^2) \ . \label{covEin}
\ee

We now attempt to repeat this calculation for the gravitational Chern-Simons part of  \eqref{eq:atmg2}. A very similar calculation yields
\be
S_{\rm cone, CS} = -\frac{i \ep}{16 \mu G_3} \int dy\;\ep^{ab} \p_a U_b  \label{CSexp}
\ee
Obtaining this expression involves an integration by parts. The validity of this procedure requires a careful understanding of the boundary conditions on the integral, as we discuss in Appendix \ref{app:coneac}.

Now we note that the explicit appearance of the metric functions $U_a(y)$ makes this expression qualitatively different from \eqref{noncovEin}:  in writing down the metric \eqref{conemetnice} we actually implicitly made a coordinate choice about how the transverse plane parametrized by $\sig^a$ rotates as we move in the $y$ direction. To be more explicit, if we perform an infinitesimal $y$-dependent rotation:
\be
\delta \sig^a = -\th(y) \ep^a_{\phantom{a}b} \sig^b
\ee
then the curl of the vector field $U_a$ transforms inhomogenously:
\be
\delta \le(\ep^{ab}\p_a U_b(y)\ri) = 4\theta'(y) \ . \label{gaugeU}
\ee
The Chern-Simons action is not invariant under this coordinate transformation. Instead it shifts by a boundary term from the endpoints of the $y$ integral:
\be
\delta S_{\rm cone, CS} = -\frac{i\ep}{4\mu G_3}\le(\theta(y_f) - \theta(y_i)\ri)~. \label{Sshift}
\ee
This is consistent with general expectations for Chern-Simons terms. However it appears to make it impossible to construct a local covariant expression analogous to \eqref{covEin}. To proceed we will need to introduce an ingredient that knows about the twisting of the coordinate system as we move along the curve. 

\begin{wrapfigure}{r}{0.25\textwidth}
\begin{center}
\includegraphics[width=0.25\textwidth]{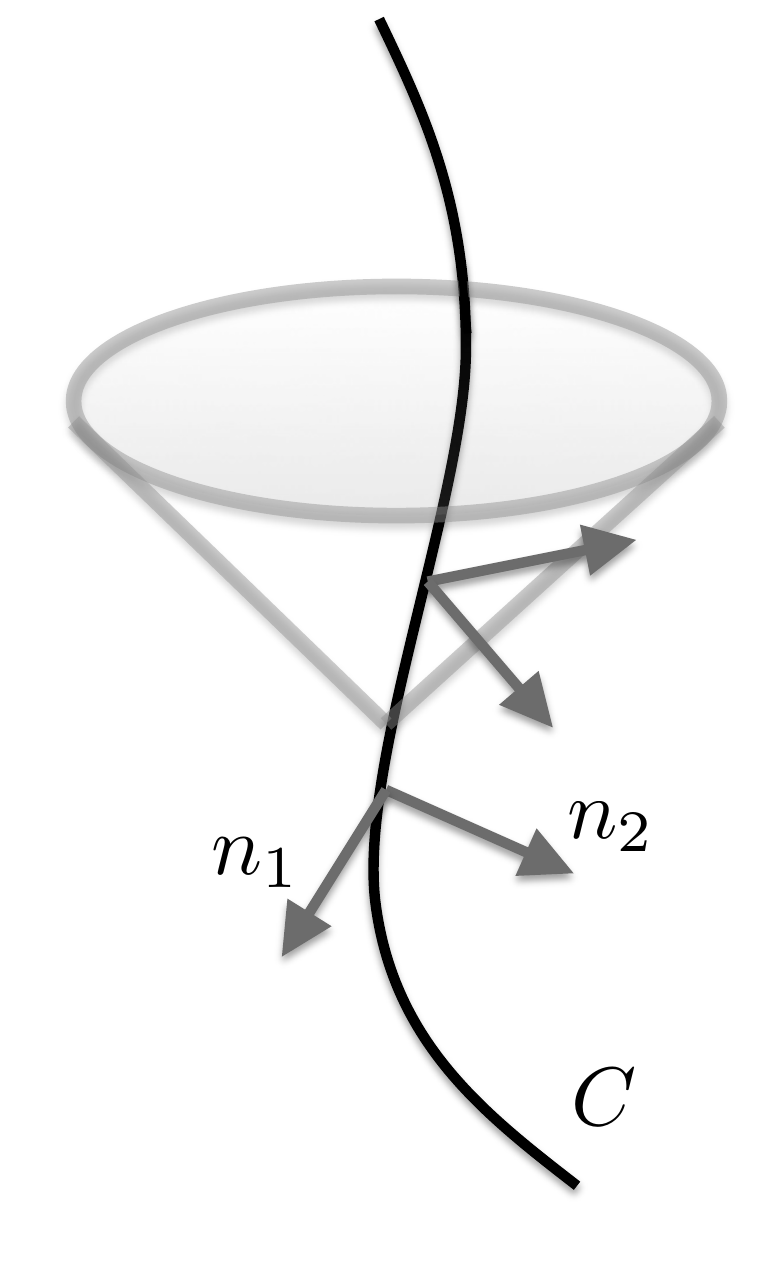}
\end{center}
\label{fig:cone}
\end{wrapfigure}

To that end, consider choosing a normal vector to the curve $n_1$. The choice of this vector fixes the other normal vector $n_2$ via $n_1 \cdot n_2 = 0$.\footnote{Up to an overall sign related to a choice of handedness.} Still there is a local $SO(2)$ freedom in this choice: for our purposes, we want these vectors to store the information of how the local coordinates change along the curve, and so we will {\it define} them with respect to the coordinate choice of the $\sig^a$, i.e:
\be
n_1 := \frac{\p}{\p\sig^1} \qquad n_2 := \frac{\p}{\p\sig^2}  \label{norms}
\ee
To lowest order in $\ep$ these vectors have unit norm, and they are both orthogonal to the tangent vector $v = \sqrt{g^{yy}}\p_{y}$. With this definition it is possible to write down a local expression which reduces to \eqref{CSexp}:
\be
S_{\rm cone, CS} = -\frac{i\ep}{4\mu G_3} \int_{C} ds\;n_2 \cdot \nabla n_1 +\ldots~, \label{spin}
\ee 
where the dots denote subleading terms in $(z,\zb)$ and $\ep$. The symbol $\nabla$ with no subscript indicates a covariant derivative along the worldline:
\be\label{def:cov}
\nabla V^\mu:= {d V^\mu\over d s}+\Gamma^{\mu}_{\lambda\rho}\frac{d X^\rho}{ds} V^\lambda \ .
\ee
We emphasize that this expression is evaluated at the tip of the cone, located in these coordinates at $\sig^a = 0$. Furthermore only the linear in $\ep$ piece of the action \eqref{spin} is expected to have a local representation along the worldline.\footnote{We briefly mention the history of the term \eqref{spin}. In three flat Euclidean dimensions this expression is known as the {\it torsion} of a curve \cite{Pressley:2010}. It is also known from the physics of anyons, where it appears along the worldline of a massive particle with a coefficient that measures the fractional spin \cite{Polyakov:184847}. It is also discussed in \cite{Witten:1988hf} in relation to the framing anomaly in Chern-Simons theory.}

The functional \eqref{spin} also {\it appears} to be covariant, but this is a delicate point, as in this derivation $n_1$ and $n_2$ were defined in terms of the base coordinate system. Essentially the anomaly has taken a bulk gravitational degree of freedom that used to be pure gauge and given it life, binding it to the worldline in the form of a normal vector. As we will explore in detail in the remainder of this paper, these normal vectors are not true dynamical degrees of freedom, but their existence and the associated subtle non-covariance of this expression is precisely what is needed to account for the {\it boundary} non-covariance associated with a gravitational anomaly.

Now using the original expression \eqref{EEdef} and taking the $\ep$ derivative, we find the total entanglement entropy in topologically massive gravity to be
\be
S_{\rm EE} =  \frac{1}{4G_3}\int_{C} ds \le(\sqrt{g_{\mu\nu} \dot{X}^{\mu}\dot{X}^{\nu}} + \frac{i}{\mu} n_2 \cdot \nabla n_1 \ri)_E~, \label{totSeuc}
\ee
where the first term is the usual Ryu-Takayanagi term and the second is the extra contribution from the Chern-Simons term. The subscript $E$ reminds us that this whole expression is evaluated in Euclidean signature. 

Let us attempt to interpret this expression in Lorentzian signature, with the path $C$ spacelike. If we try to analytically continue the complex coordinates $(z,\bar{z})$ via $z = x - t, \bar{z} = x + t$, then we see from \eqref{norms} that we can obtain two real Lorentzian normal vectors $(n, \tilde{n})$ via
\be
n :=i n_1 =  \p_t \qquad \tilde{n} := n_2 = \p_x ~.   \label{lorN}
\ee
We choose a notation that is no longer symmetric with respect to the two normal vectors, as one of them is now timelike $(n)$ and the other spacelike $(\tilde n)$. Expressing \eqref{totSeuc} in terms of these vectors we now find the following expression for the entanglement entropy:
\be
\boxed{S_{\rm EE} = \frac{1}{4G_3}\int_{C} ds \le(\sqrt{g_{\mu\nu} \dot{X}^{\mu}\dot{X}^{\nu}} + \frac{1}{\mu} \tilde{n} \cdot \nabla n \ri) \label{sEElor}}
\ee
where everything is now evaluated in Lorentzian signature. 

This expression is one of the main results of this paper. It is manifestly real. However unlike the steps leading to \eqref{totSeuc}, this analytic continuation cannot in general be justified, as the $\sig^a$ coordinate system used above is ill-defined away from a vicinity of $C$: certainly there is generally no $U(1)$ isometry along which we can continue the bulk spacetime. Rather the relation between \eqref{totSeuc} and this formula is equivalent to the relation between the (justified) Ryu-Takayanagi formula for static spacetimes and its (yet unproven) covariant generalization \cite{Hubeny:2007xt}. Nevertheless, we will use it for the rest of the paper and find physically sensible results. 

Next, we note that this expression should be evaluated on a worldline $C$. As it turns out, the correct worldline is determined by extremizing the functional \eqref{totSeuc}. As emphasized in \cite{Dong:2013qoa, Camps:2013zua}, the justification of this statement is in principle a different question than the determination of the functional itself. We present two routes to its justification in Appendix \ref{app:back}. We show that the consistency of the TMG equations near a spacetime with a conical defect constrains the worldline of the defect in a way that requires \eqref{totSeuc} to be extremized, and we show that viewing \eqref{totSeuc} as a {\it source} to the TMG equations creates the desired conical defect. We now move on to discuss the physical content of the resulting equations of motion. 

\subsection{Spinning particles from an action principle} \label{sec:spinact}

Let us take a step back and recapitulate what we expect the cone action to describe in the bulk.
As articulated at the start of this Section, 
the action \eqref{sEElor} should capture the physics of a heavy particle with mass $\fm$ and a continuously tunable spin $\fs$ determined by the CFT twist operator quantum numbers \eqref{twdims}: in other words, we expect an {\it anyon} in curved space. In this section we will flesh out this interpretation, and to make it evident  in what follows we write the entanglement functional \eqref{sEElor} as
\be
S_{\rm EE} = \int_{C} ds \le(\fm \sqrt{g_{\mu\nu} \dot{X}^{\mu}\dot{X}^{\nu}} + \fs\;\tilde{n} \cdot \nabla n \ri), \label{sEElor2}
\ee
with $\fm$ and $\fs$ real constants. 

The variational principle for \eqref{sEElor2} should include variations with respect to both the particle position $X^{\mu}(s)$ and the normal vectors $(n, \tilde{n})$. However, our construction requires that $(n, \tilde{n})$ remain normal to the curve , and so if we are to vary them then \eqref{sEElor2} must be supplemented   with the following constraint action:
\be\label{eq:con} 
S_{\rm constraints}=\int_{C} ds \left[\lambda_1 n \cdot \tilde{n} + \lambda_2 n \cdot v +\lambda_3 \tilde{n} \cdot v +\lambda_4 (n^2+1) +\lambda_5(\tn^2-1)\right]~,
\ee
where the $\lam_i$ are five Lagrange multipliers that guarantee that $n$ and $\tn$ are normalized, mutually orthogonal, and perpendicular to the worldline:
\be\label{eq:ort}
n \cdot v =0 ~,\quad \tn \cdot v =0 ~,\quad n \cdot \tn =0, \quad n^2 = -1, \quad \tn^2 = 1 ~,
\ee
which also enforce that $n$ is timelike and $\tilde n$ is a spacelike vector. These are the constraints that allow us to write  \eqref{sEElor} in the first place. Here $v^{\mu}$ is the velocity vector:
\be\label{def:x}
v^\mu:= {d X^\mu\over ds} = \dot X^\mu~.
\ee
We choose the affine parameter $s$ of this path to be the proper length along the path and hence we have $v^\mu v_\mu=1$. The total action for the spinning particle is then
\be
S_{\rm probe} = \int_{C} ds \le(\fm \sqrt{g_{\mu\nu} \dot{X}^{\mu}\dot{X}^{\nu}} + \fs\;\tilde{n} \cdot \nabla n \ri) + S_{\rm constraints}~. \label{eq:actp3}
\ee

In addition to the particle trajectory $X^{\mu}(s)$, it {\it appears} that we now have a dynamical degree of freedom corresponding to the rotation of the normal frame along the worldline. Indeed we have three independent components for each of $n$ and $\tilde{n}$, and five constraints associated with \eqref{eq:con}, leaving a single degree of freedom. However as it turns out this is not a true degree of freedom, as the action is only sensitive to its variation up to boundary terms.
To see this, consider the equation of motion arising from variation of $S_{\rm probe}$ with respect to $n$: 
\be\label{eq:en1}
-\mathfrak{s}\nabla \tn^\mu +\lambda_1 \tn^{\mu}+\lambda_2 v^\mu +2\lambda_4 n^{\mu} = 0~.
\ee
Contracting with $\tn^{\mu}$, $v^\mu$ and $n^\mu$ respectively results in
\be\label{eq:l1}
\lam_1 = \fs\;\tn \cdot \nabla \tn \qquad \lambda_2= \mathfrak{s}\, v\cdot\tn ~,\quad 2\lambda_4= -\mathfrak{s}\,n \cdot\nabla \tn~.
\ee
The same analysis for $\delta_{\tn}S_{\rm probe}=0$ gives 
\be\label{eq:en2}
\mathfrak{s}\nabla \tn_{\mu}+\lambda_1 n^{\mu}+\lambda_3 v^\mu +2\lambda_5 \tn^{\mu} = 0,
\ee
and again the appropriate contractions yield
\be\label{eq:l2}
\lam_1 = \fs\,n \cdot \nabla n \qquad \lambda_3= -\mathfrak{s}\, v\cdot\nabla n ~,\quad 2\lambda_5=-\mathfrak{s}\,\tn \cdot\nabla n~.
\ee
\eqref{eq:l1} and \eqref{eq:l2} are six equations, of which five may be satisfied by appropriate choice of the $\lam_i$. The equation that remains comes from the fact that $\lam_1$ appears in both sets of equations, and is:
\be
\tn \cdot \nabla \tn = n \cdot \nabla n~.
\ee
However this is not a dynamical equation: as $-n^2 = \tn^2 = 1$, both sides of this equation are identically zero and this is an identity. Thus $(n, \tn)$ do not have a dynamical equation of motion: equivalently, $S_{\rm probe}$ is insensitive to small variations of the normal frame along the trajectory, up to boundary terms. As we will see, these boundary terms will be very important to us. 

A more tedious task is to vary \eqref{eq:actp3} with respect to $X^\mu(s)$; details can be found in Appendix \ref{app:p3}. This variation, however, is not trivial and gives rise to 
\bea\label{eq:VV}
\nabla\left[\mathfrak{m} v^\mu + v_\rho \nabla s^{\mu\rho}\right]=-{1\over 2}v^\nu s^{\rho\sigma}R^\mu_{~\nu\rho\sigma}~,
\eea
where we define the {\it spin tensor} $s^{\mu\nu}$ to be 
\be\label{def:s}
s^{\mu\nu}= \mathfrak{s} \left(n^\mu \tn ^\nu -\tn^{\mu} n^\nu\right)  ~,
\ee
These equations are known as the Mathisson-Papapetrou-Dixon (MPD) equations, and describe the motion of spinning particles in classical general relativity \cite{Mathisson:1937zz,Papapetrou:1951pa,Dixon:1970zza}.\footnote{See as well \cite{Leclerc:2005yy, Porto:2005ac,Vasilic:2007wp,Vasilic:2008gt,Armas:2013hsa} for a more modern treatment and further references.} In $(2+1)$ dimensions they follow from the simple and geometric action \eqref{eq:actp3}. 

Next, using \eqref{eq:ort} we see that the spin tensor may be written in terms of the velocity as 
\be\label{eq:sv}
s^{\mu\nu}=-\mathfrak{s} \epsilon^{\mu\nu\lambda}v_\lambda~,
\ee
indicating that the spin tensor is just the volume form perpendicular to the worldline and does not actually carry any extra degrees of freedom.\footnote{In higher dimensions such a rewriting is not possible: $s^{\mu\nu}$ there actually carries information about the direction of the particle's spin. In that case the MPD equations also include a relation for the evolution of the spin tensor, 
\be\label{eq:spin}
\nabla s^{\alpha\beta}+v^\alpha v_\mu \nabla s^{\beta\mu} -v^\beta v_\mu \nabla s^{\alpha\mu}=0~,  
\ee
which is an identity on \eqref{eq:sv}.} Note also that the normal vectors themselves have vanished from the equations of motion: they are required only to set up the action principle.  

Thus to compute entanglement entropy in topologically massive gravity we should solve the MPD equations \eqref{eq:VV} using the values of $
\fm$ and $\fs$ inherited from \eqref{sEElor}:
\be\label{eq:msmpd}
\fm = \frac{1}{4 G_3}~, \quad \fs = \frac{1}{4 \mu G_3}~,
\ee
 and then evaluate the action \eqref{sEElor} on the resulting solution. 

\section{Applications}\label{sec:results}

In this section we finally apply our prescription to the actual computation of holographic entanglement entropy on various backgrounds of interest. We will compare with the CFT results found in Section \ref{sec:cft}. Recall that the prescription is as follows: the entanglement functional \eqref{sEElor} is 
\be
S_{\rm EE} = \frac{1}{4G_3}\int_{C} ds \le(\sqrt{g_{\mu\nu} \dot{X}^{\mu}\dot{X}^{\nu}} + \frac{1}{\mu} \tilde{n} \cdot \nabla n \ri)~. \label{repEE}
\ee
It should be evaluated on a curve $C$ in spacetime which is found by extremization of the functional itself, which results in the Mathisson-Papapetrou-Dixon equations \eqref{eq:VV} for the motion of a spinning particle of mass $\fm$ and spin $\fs$ given by \eqref{eq:msmpd}. 
 
We will limit our discussion to spacetimes that are locally AdS$_3$. This greatly simplifies the analysis, as here the contraction of the Riemann tensor with $s_{\mu\nu}v_{\rho}$ vanishes.\footnote{This can be shown by writing the Riemann tensor in terms of the metric using \eqref{eq:RR}.} The MPD equations then reduce to
\be\label{eq:p33d}
\nabla\left[\mathfrak{m} v^\mu -  \mathfrak{s} \epsilon^{\mu\nu\lambda}v_\nu \nabla v_\lambda \right]=0~,
\ee
and furthermore, it is clear that a geodesic
\be
\nabla v^\mu =0~,
\ee
is a solution to  \eqref{eq:p33d}. Evidently anyons moving on maximally symmetric spaces follow geodesics.\footnote{Note that \eqref{eq:p33d} {\it does} also have other solutions, as it is a higher-derivative equation of motion: we believe (without proof) that they will have greater actions, and in this work we will not investigate them.}  

So all we need to do is evaluate the second term in the action \eqref{repEE} on a geodesic. We call this term $S_{\rm anom}$. For this purpose it will be helpful to rewrite the action in terms of a single normal vector $n$ using $\tn^{\mu} = \ep^{\mu\nu\rho} v_{\nu} n_{\rho}$ to find 
\be
S_{\rm anom} = \frac{1}{4 G_3 \mu} \int_C ds\; \ep_{\mu\nu\rho}v^{\mu} n^{\nu}(\nabla n^{\rho}) \ . \label{torsion}
\ee 
We note that although the equations of motion can be entirely formulated in terms of the trajectory $X^{\mu}(s)$, the on-shell action itself depends on extra data, namely boundary data of the normal vectors $n(s)$, which for the moment we parametrize as follows:
\be
n(s_i) = n_i ~,\quad n(s_f) = n_f, \label{nbc}
\ee
where $n_i$ and $n_f$ are normal vectors defined in the CFT. As discussed in Section \ref{sec:spinact}, the action is actually insensitive to smooth variations of $n(s)$ that leave $n_i$ and $n_f$ unchanged: it measures only the twist of $n_f$ relative to $n_i$. 

\begin{figure}[h]
\includegraphics[width=0.45\textwidth]{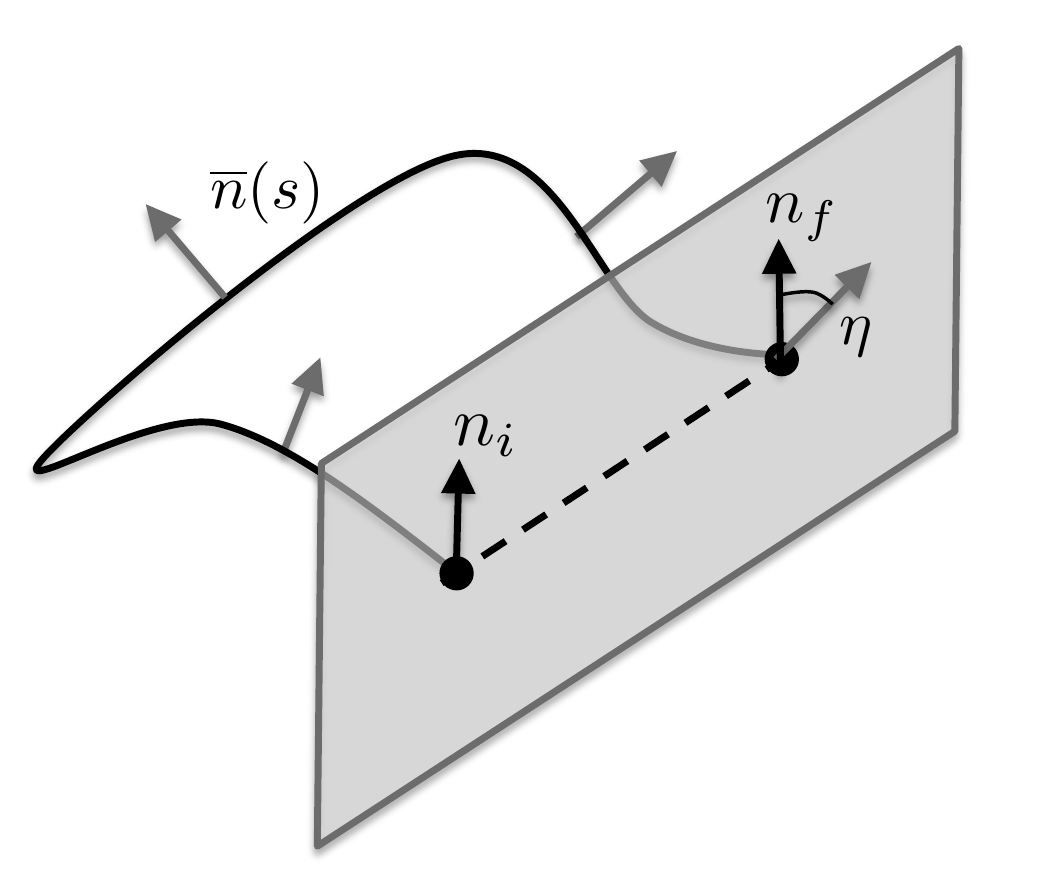}
\caption{Propagating $n_i$ into $\overline{n}(s)$ (denoted in gray) along the curve by solving parallel transport equation \eqref{partrans}. In general $\overline{n}(s_f)$ will be related to $n_f$ by a boost with angle $\eta$. }
\label{fig:nbar}
\end{figure}

What does this actually mean in curved space? We require a way to compare $n_f$ to $n_i$. To that end, consider parallel transporting $n_i$ along the curve, i.e. consider the solution to the first-order parallel-transport\footnote{This construction only applies if the curve is a geodesic; in a more general case one should replace parallel transport with Fermi-Walker transport (see e.g. \cite{Poisson:2003nc}) to guarantee that $\overline{n}(s)$ remains normal to the curve, but the intuition is still valid.} equation
\be
\nabla \overline{n}(s) = 0~, \quad \overline{n}(s_i) = n_i \ . \label{partrans}
\ee
This has a unique solution all along the curve, as illustrated in Figure \ref{fig:nbar}. However $\overline{n}(s)$ is {\it not} equal to $n(s)$, as it will in general not satisfy the right boundary condition at $s_f$. Instead at the endpoint it will be related to $n_f$ via an $SO(1,1)$ transformation:
\be
\overline{n}(s_f)^a = \Lam(\eta)^{a}_{\phantom{a}b} n_f^b \label{lorrel} \ .
\ee
As we now show, the integral \eqref{torsion} measures the rapidity $\eta$ of this Lorentz boost $\Lam$. This is the generalization of the idea of ``twisting'' to curved space (and Lorentzian signature). 

To prove this assertion, consider a vector $n(s)$ that actually satisfies the boundary condition \eqref{nbc}. Consider also a parallel transported normal frame along the path given by two vectors $(q^{\mu}, \tilde{q}^{\mu})$, $\nabla q = \nabla \tilde{q} = 0$. We can expand $n(s)$ in terms of $q$ and $\tilde{q}$: 
\be
n(s) = \cosh(\eta(s)) q(s) + \sinh(\eta(s)) \tilde{q}(s) \label{nexp}
\ee
The integral \eqref{torsion} now reduces to the total derivative
\be
S_{\rm anom} = \frac{1}{4G_3 \mu} \int ds\, \dot\eta(s) = \frac{1}{4 G_3\mu} \le(\eta(s_f) - \eta(s_i)\ri) \ .
\ee
On the other hand, by its definition \eqref{partrans} we have
\be
\overline{n}(s) = \cosh(\eta(s_i)) q(s) + \sinh(\eta(s_i)) \tilde{q}(s), \label{ovn}
\ee
i.e. its $s$-dependence comes entirely from the parallel transport of the normal frame and not from any rotation of the frame itself. Comparing \eqref{ovn} to \eqref{nexp} we conclude that $S_{\rm anom}$ measures the twist. \eqref{nexp} alone is sufficient to determine $S_{\rm anom}$: in fact, once we find $q$ and $\tilde{q}$ we see that the total twist can be conveniently written as:
\be
S_{\rm anom} = \frac{1}{4G_3 \mu}\log\le(\frac{q(s_f) \cdot n_f - \tq(s_f) \cdot n_f}{q(s_i) \cdot n_i - \tq(s_i) \cdot n_i}\ri),\label{covanswer}
\ee 
Note that despite the topological character of $S_{\rm anom}$ with respect to variations of $n$, it depends (through the parallel transport equation) continuously on parameters of the bulk metric and thus on the state of the field theory.  

Finally, we discuss the explicit choice of $n_i$ and $n_f$. Recall from the original definition of the normal frame back in \eqref{norms} that the normal vectors should be defined with respect to the coordinate system used to parameterize the bulk. At the endpoints of the interval the coordinate system in the bulk coincides with that used to define the CFT, and so \eqref{lorN} instructs us to take:
\be
n_i = n_f = \le(\p_t\ri)_{\rm CFT}~. \label{CFTbc}
\ee
Here (and in the following subsections) we define $\le(\p_t\ri)_{\rm CFT}$ as a vector that points in the time direction at the boundary, however it is normalized to ensure that $n^2=-1$ with respect to the bulk metric. The direction along $(\p_t)_{\rm CFT}$ is interpreted  as the field theory time coordinate that is used to define the vacuum of the theory. The fact that the action $S_{\rm anom}$ is insensitive to smooth variations of $n(s)$ in the interior follows from the fact that the physics is coordinate-invariant in the bulk. However it is a nontrivial fact that the entanglement entropy now depends in a subtle way on this choice of time coordinate on the {\it boundary}: as we saw in Section 2, this is precisely what one expects from a theory with a gravitational anomaly.

We turn now to some explicit computations.

\subsection{Poincar\'e AdS}
\begin{figure}[h]
\includegraphics[width=0.8\textwidth]{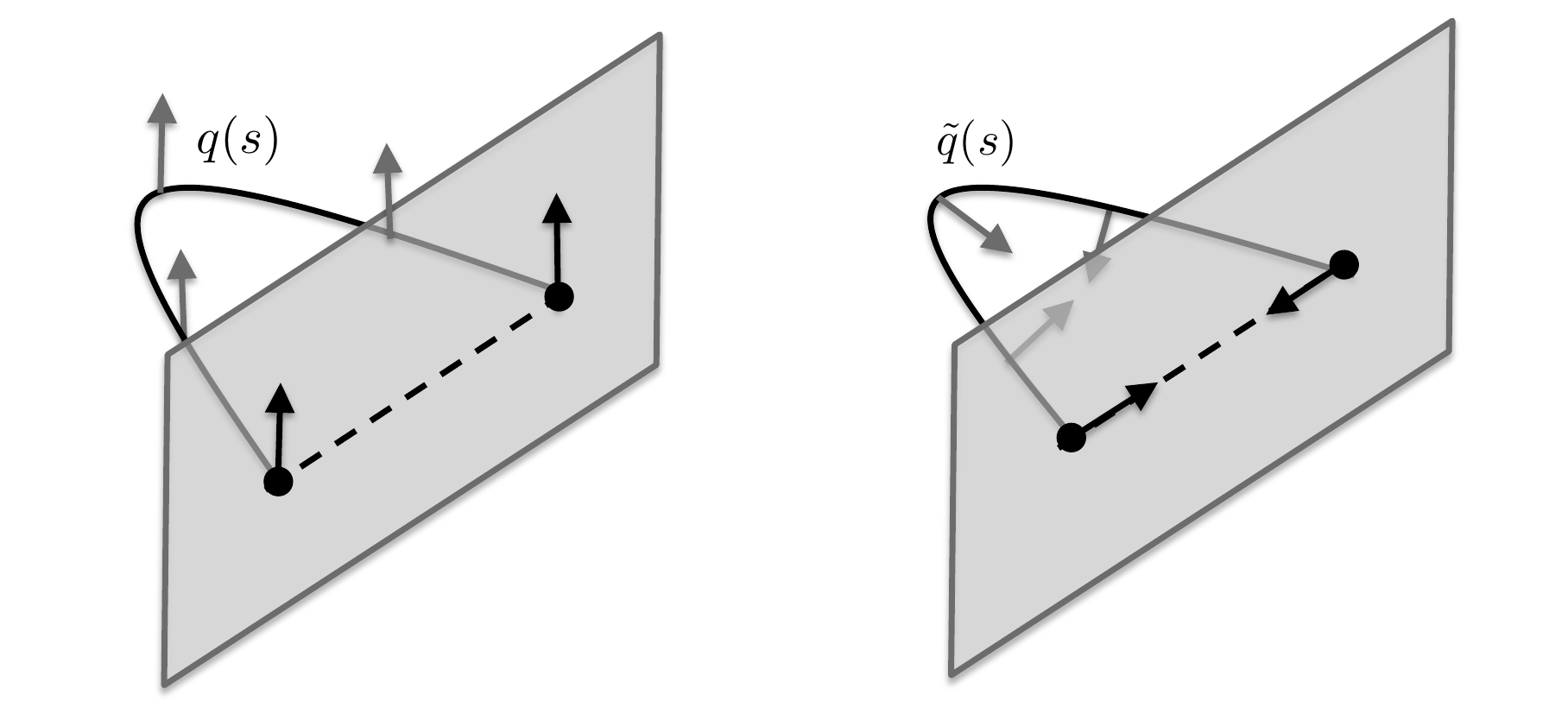}
\caption{Evolution of normal frame vectors $q(s)$ and $\tilde{q}(s)$ for an interval at rest in  Poincar\'e AdS$_3$. }
\label{fig:qs}
\end{figure}

We first study AdS$_3$ in Poincar\'e coordinates, corresponding to the vacuum of the CFT$_2$ defined on a line. The metric is
\be
ds^2={\ell^2\over u^2}(-dt^2 + dx^2 +du^2)~. \label{AdSPoinc}
\ee
The boundary is at $u = 0$. We first consider an interval {\it at rest}: that is, we want to compute the entanglement entropy of an interval of length $R$ in the vacuum, stretching from $(x,t) = \le(-\frac{R}{2},0\ri)$ to $\le(\frac{R}{2},0\ri)$. The geodesic equation admits the following solution:
\be
{u^2+ x^2 =\frac{R^2}{4}}~,
\ee
The tangent vector and a parallel transported normal frame is (using $(x,t,u)$):
 \be
 v^\mu={2 u\over R\ell}(u,0,-x)~, \quad q^\mu={1\over \ell}(0,u,0)~, \quad \tilde{q}^{\mu} = -{2 u\over R\ell}(x,0,u)~. \label{vnorms}
 \ee
 By construction, notice that
 \be
v^2= \tilde{q}^2=1 ~,\quad q^2=-1 ~,\quad q \cdot \tilde{q} =v\cdot q =v\cdot \tilde{q} =0
 \ee
 and 
 \be
\nabla v = 0\ , \quad \nabla q=0\ , \quad \nabla \tilde{q} = 0 \ .
 \ee
 Furthermore, note that
 \be
 q(s_i) = q(s_f)= \le(\p_t\ri)_{\rm CFT} ~,\qquad \tilde{q}(s_i) = -\tilde{q}(s_f)= \le(\p_x\ri)_{\rm CFT}~, \label{poinc}
 \ee
where $\le(\p_t\ri)_{\rm CFT}$ points in the time direction at the boundary and it is normalized such that $q^2=-1$ with respect to the bulk metric; a similar definition applies to $\le(\p_x\ri)_{\rm CFT}$. The normal vector perpendicular to the boundary interval does not change under parallel transport, whereas the vector that is parallel to the boundary interval rotates through $\pi$: this is intuitively obvious from Figure \ref{fig:qs}.

It is simple to evaluate \eqref{nexp}: we see that $q(s)$ already satisfies the boundary conditions \eqref{CFTbc}. This means that a vector pointing in the time direction {\it still} points in the time direction even after being parallel transported to the other end of the curve; there is no twisting. So \eqref{nexp} reduces to
 \be
 n(s) = q(s) \quad \to \quad \eta(s) = 0~.
 \ee
 Thus the spin term in \eqref{repEE} does not contribute, the entanglement entropy is simply given by the proper distance, and we find the usual CFT$_2$ expression \eqref{gdee}:
 \be
 S_{EE} = \frac{c_L + c_R}{6} \log\le(\frac{R}{\varepsilon}\ri)~.
 \ee

\subsection{Poincar\'e AdS, boosted interval}\label{sec:adsbi}
Let us now consider the entanglement in a {\it boosted} interval $(x',t')$ relative to $(x,t)$ in \eqref{AdSPoinc}. In the $(x,t)$ coordinates we have a spacelike line stretching from 
\be\label{eq:intadsb}
 -\frac{R}{2}(\cosh \ka, \sinh \ka)\quad {\rm to} \quad \frac{R}{2}(\cosh \ka, \sinh \ka),
\ee
with $\ka$ a boost parameter. In a Lorentz-invariant theory the entanglement entropy would be insensitive to $\ka$. Much of the calculation can be taken over from above: we can obtain the appropriate $q$, $\tilde{q}$ simply by boosting \eqref{vnorms}. Explicitly, we find
\be\label{eq:qqads}
q^{\mu} = \frac{1}{\ell}\le(u \sinh \ka, u \cosh \ka, 0\ri)~, \quad \tq^{\mu} = -\frac{2u}{R \ell}\le(x \cosh \ka, x \sinh \ka, u\ri) \ .
\ee
Importantly, however, $n_i = n_f$ still points purely in the time direction of the CFT, and thus now has a component directed along the boundary interval. Evaluating \eqref{covanswer} we now find
\be
S_{\rm anom} = \frac{1}{2 G_3 \mu} \ka~,
\ee
and the total entanglement entropy is
\be
S_{EE} = \frac{c_L + c_R}{6} \log\le(\frac{R}{\varepsilon}\ri) - \frac{c_L - c_R}{6} \ka, \label{eeboost}
\ee 
This matches the Lorentzian CFT result \eqref{eevac}: the entanglement entropy depends on the boost of the interval relative to the vacuum.

\subsection{Rotating BTZ black hole}
We turn now to the rotating BTZ black hole, which has metric
\be\label{eq:metricbtz}
\frac{ds^2}{\ell^2} = -\frac{(r^2 - r_+^2)(r^2 - r_-^2)}{r^2} d\tau^2 + \frac{r^2}{(r^2 - r_+^2)(r^2 - r_-^2)} dr^2 + r^2 \le(d\phi + \frac{r_+ r_-}{r^2}d\tau\ri)^2 \ .
\ee
This is dual to a field theory state with unequal left and right-moving temperatures, cf. \eqref{tempsss}:
\be
\beta_L = \frac{2\pi \ell}{r_+ - r_-} \qquad \beta_R = \frac{2\pi \ell}{r_+ + r_-} \ .
\ee
The boundary CFT lives on a flat space with metric:
\be
ds^2_{\rm CFT} = -d\tau^2 + d\phi^2 ~.\label{BTZbdy}
\ee 
We will assume that the $\phi$ coordinate is noncompact: thus technically this is not a rotating black hole but rather a boosted black brane, i.e. a rotating BTZ black hole at high temperature. We will consider a boundary interval of length $R$ along the $\phi$ direction at time $\tau = 0$. The boundary conditions on the normal vector are defined in terms of the time coordinate appropriate to the black hole,
\be
n_i = n_f = \le(\p_{\tau}\ri)_{\rm CFT} \ . \label{BTZns}
\ee
In this case as we parallel transport $\overline{n}$ into the bulk, we expect it to be dragged by the boosted black hole horizon, picking up a nontrivial boost.  

Rather than directly solve the differential equation \eqref{partrans} to compute this boost, we can simplify our computation by using the fact that the BTZ black hole is locally equivalent to AdS$_3$. The explicit mapping to AdS$_3$ Poincar\'e coordinates \eqref{AdSPoinc} is:
\be
x \pm t = \sqrt{\frac{r^2 - r_+^2}{r^2 - r_-^2}} e^{2\pi (\phi \pm \tau)/\beta_{R,L}}~, \qquad u = \sqrt{\frac{r_+^2 - r_-^2}{r^2 - r_-^2}} e^{(\phi r_+ + \tau r_-)/\ell}  ~.\label{poincBTZ}
\ee
We may now essentially take over the results from the previous section. We use \eqref{poincBTZ} to map the endpoints of the interval $(\phi, \tau)_{1,2} \equiv \le\{ \le(-\frac{R}{2}, 0\ri), \le(\frac{R}{2},0\ri)\ri\}$ to the Poincar\'e coordinates $(x,t)_{1,2}$. Denote by $R_P$ the length of the boundary interval in the Poincar\'e conformal frame, i.e.
\be
R_P := \sqrt{(x_2 - x_1)^2 - (t_2 - t_1)^2} \ .
\ee
It is now convenient to construct a 2d normal vector $p^i$ that is parallel to the boundary interval in Poincar\'e coordinates, i.e in components
\be
p^i = \frac{1}{R_P}\le(x_2 - x_1, t_2 - t_2\ri)
\ee
where the index $i$ runs over $x$ and $t$. The $q$'s in \eqref{eq:qqads} can now be written (again in Poincar\'e coordinates) as
\be
q^{\mu} = \frac{u}{\ell}\le(\ep^i_{\phantom{i}j} p^j, 0\ri) \qquad \tilde{q}^{\mu} =- \frac{2 u}{\ell R_P}\le(\lam p^i, u\ri) \qquad \lam^2 + u^2 = \frac{R_P^2}{4} \label{qqBTZ}
\ee
Here $\lam$ parametrizes movement along the curve, essentially playing the role that $x$ played in \eqref{eq:qqads}, starting out at $\lam(s_i) = -\frac{R_P}{2}$ at the left endpoint and ending at $\lam(s_f)=  \frac{R_P}{2}$ at the right endpoint. Note that \eqref{qqBTZ} simply states that $q$ remains perpendicular to the boundary interval while $\tilde{q}$ has a component directed along it that switches sign as we move from one endpoint to the other.

It is now straightforward to convert the $q$'s back to BTZ coordinates and compute the inner products in \eqref{covanswer} using \eqref{BTZns}. We find after some algebra
\be
S_{\rm anom} = \frac{1}{4 G_3 \mu} \log\le(\frac{\sinh\le(\frac{\pi R}{\beta_R}\ri) \beta_R}{\sinh\le(\frac{\pi R}{\beta_L}\ri) \beta_L}\ri)~,
\ee
and thus the total entanglement entropy is 
\be
S_{\rm EE} = \frac{c_R + c_L}{12} \log\le(\frac{\beta_L \beta_R}{\pi^2 \varepsilon^2} \sinh\le(\frac{\pi R}{\beta_R}\ri) \sinh\le(\frac{\pi R}{\beta_L}\ri)\ri) + \frac{c_R - c_L }{12} \log\le(\frac{\sinh\le(\frac{\pi R}{\beta_R}\ri) \beta_R}{\sinh\le(\frac{\pi R}{\beta_L}\ri) \beta_L}\ri), \label{entEE}
\ee
where we have added back the ordinary proper distance piece, first computed holographically in \cite{Hubeny:2007xt}. This agrees with the result computed from field theory in \eqref{eerot}. Note that the structure of the second term is precisely correct to allow the answer to be written as a sum of separate left and right moving contributions. 

\subsection{Thermal entropies}
The evaluation of the thermal entropy of a black hole in topologically massive gravity has been studied previously \cite{Solodukhin:2005ah,Tachikawa:2006sz} and we would like to discuss the connection with our formalism. 

Note that if we are evaluating the action of a closed loop wrapping a black hole horizon, there is no longer a choice of boundary conditions on the normal vectors. Any choice of normal vectors will give the same answer, provided that it is single-valued around the circle. A simple choice is just to take the normal frame to be constant in any convenient coordinate system:
\be
\frac{d}{ds}n^{\mu} = \frac{d}{ds} \tn^{\nu} = 0 \label{constN}
\ee
in which case the spin term in \eqref{repEE} simply becomes
\be\label{eq:Wald}
S_{\rm anom} = \frac{1}{4 G_3 \mu} \oint_{\mathcal{H}} ds\;\Ga^{\mu}_{\al\beta} n^{\beta}\tn_{\mu}v^\al,
\ee 
with $\Ga^{\mu}_{\al\beta}$ the usual affine connection. This is equivalent to the expression in \cite{Solodukhin:2005ah,Kraus:2005vz,Tachikawa:2006sz}. It is gauge-invariant under coordinate changes that are single-valued around the circle. From our point of view it is measuring the boost acquired by a normal vector if it is parallel transported around the horizon and then compared to itself. Note that \eqref{constN} does not mean that the vector is {\it covariantly} constant; the distinction between these two notions is precisely what the spin term measures.

For concreteness, we evaluate \eqref{eq:Wald} for the BTZ black hole. Using \eqref{eq:metricbtz} and \eqref{constN} we have
\be
n= \frac{\sqrt{g_{rr}}}{\ell^2}(\partial_t-{r_+r_-\over r^2}\partial_\phi)~,\quad \tn =-{1\over \sqrt{g_{rr}}} \partial_r~,\quad  v={1\over \ell r}\partial_\phi~.
\ee
Strictly speaking these vectors are evaluated at $r=r_+$, and hence are singular; but for the purpose of computing \eqref{eq:Wald} this pathology drops out. We find  
\bea
S_{\rm anom} = \frac{\pi r_-}{2 G_3 \ell \mu}  \ .
\eea
After adding the contribution from the area law, this exactly reproduces \eqref{cardy}. Note that the entropy density extracted from this expression is consistent with that arising from the $R \to \infty$ limit of \eqref{entEE}.  

\section{Discussion} \label{sec:conc}
We studied holographic entanglement entropy in AdS$_3$/CFT$_2$ in the presence of a gravitational anomaly. Our main result can be easily stated. The gravitational anomaly introduces a non-trivial dependence on the choice of coordinates when evaluating  entanglement entropy. This data  is transmitted into the bulk by broadening the Ryu-Takayanagi minimal worldline into a {\it ribbon}, i.e. a bulk worldline together with a normal vector.

The resulting contribution of the anomaly to the entanglement entropy is pleasantly geometric:
\be
S_{\rm anom} = \frac{1}{4 G_3 \mu} \int_{C} ds \le( \tilde{n} \cdot \nabla n \ri), \label{finalS}
\ee
where $\tn$ and $n$ define a normal frame. This expression measures the net twist of this normal frame along the worldline. 
In the bulk this action can be interpreted as describing a spinning particle with a continuously tunable spin -- an {\it anyon} -- in $(2+1)$ dimensions. In particular, we showed that the minimization of the total action reproduces the Mathisson-Papapetrou-Dixon equations for the motion of a spinning particle in general relativity. Thus our work can also be viewed as the construction of the worldline action describing anyons in curved space. In the context of entanglement entropy, this anyon is the holographic avatar of the CFT twist operator. Finally, we computed the entanglement entropy for a single interval on various simple spacetimes to illustrate the formalism, demonstrating agreement with a field theory analysis of CFTs with $c_L \neq c_R$. 

Our prescription admits a natural extension to multiple intervals. We saw in Section \ref{sec:cft} that for holographic CFTs, the multiple interval entanglement entropy on $\mathbb{C}$ or $S^1\times \mathbb{R}$ is universal to leading order in large total central charge. It is of course precisely  this regime, and for this class of CFTs, to which bulk calculations in TMG apply. Thus it is reasonable to conjecture the following: the bulk object that computes $N$-interval entanglement entropy in holographic CFTs with $c_L\neq c_R$ is a sum over $N$ copies of our functional \eqref{sEElor}, minimized over all pairings of boundary points allowed by the homology constraint of Ryu and Takayanagi. This is precisely analogous to the original Ryu-Takayanagi prescription for multiple intervals, only now the functional is extended to include \eqref{finalS} for each geodesic. In Poincar\'e AdS, we showed that the anomalous term \eqref{finalS} does not contribute (for spatial intervals); however, it does in the planar BTZ case.

There are some natural directions for future research: 
\begin{enumerate}

\item There are other solutions to topologically massive gravity, such as warped AdS. These geometries are thought to be dual to somewhat mysterious field theories -- {\it warped CFTs} \cite{Hofman:2011zj,Detournay:2012pc}. Our construction opens the door to computing entanglement entropy in these geometries. Doing this in the field theory itself appears challenging: the tools used in Section \ref{sec:cft} have not been generalized to warped CFTs, and there are subtleties in calculating entropies related to the choice of ensemble. There are also no known microscopic examples of warped CFTs, away from certain limiting cases.\footnote{See \cite{Compere:2013aya} for a candidate example.} We take the view that an application of our construction to warped AdS may provide the first, albeit holographic, calculation of entanglement entropy in a theory with warped conformal symmetry. Regardless of considerations of warped CFTs, it seems interesting in its own right to apply our formalism to warped AdS.\footnote{It should be noted that there are also warped AdS solutions of Einstein gravity coupled to matter, {\it without} a gravitational Chern-Simons term. Holographic entanglement entropy was studied in that context in \cite{Anninos:2013nja} using a perturbative scheme that permitted application of the Ryu-Takayanagi formula. Their results favor the hypothesis that those geometries describe states in an ordinary Virasoro CFT. The outcome of holographic entanglement calculations in warped AdS solutions of TMG is likely to be different.} This will be slightly more involved than the simple applications presented in Section \ref{sec:results} (which heavily used the fact that the spinning particle equations on a maximally symmetric space admit geodesic solutions) and will likely actually require the solution of differential equations.

\item In a similar spirit as the previous point, one could consider TMG with asymptotically flat boundary conditions at null infinity. In that case the asymptotic symmetries consist of the so-called BMS$_3$ algebra, which can be obtained as an ultra-relativistic limit of the AdS$_3$ Virasoro algebra \cite{Barnich:2006av}. The corresponding phase space contains the flat limit of the BTZ black holes, which were recognized as cosmological solutions with non-trivial Bekenstein-Hawking entropy \cite{Barnich:2012xq, Bagchi:2012xr}. A challenge is to determine whether a BMS$_3$-invariant field theory living at null infinity could in some sense be dual to flat space in (2+1)-dimensions. Again, little is known about these putative field theories besides specific examples (such as the one presented in \cite{Barnich:2013yka}, which is to flat space what the Liouville CFT is to 3d gravity with a negative cosmological constant) and general properties such as the form of correlation functions \cite{Bagchi:2009my, Bagchi:2009ca} or thermal entropies \cite{Barnich:2012xq, Bagchi:2012xr}. An interesting limit of the above set up is when the bulk theory consists only of the gravitational Chern-Simons piece. In that case, the asymptotic symmetry algebra reduces to a chiral copy of a Virasoro algebra, suggesting the existence of a flat version of the $AdS_3$ chiral gravity story \cite{Bagchi:2012yk}. It would be interesting to study entanglement entropy in these theories in order to further probe the nature of the putative dual theory.

\item In any dimension, the bulk object that computes entanglement entropy is codimension $2$ and admits two normal vectors $\tn$ and $n$. Morever, action principles for {\it spinning} membranes have been constructed  (and studied) in e.g. \cite{Armas:2013hsa}.  Thus some features of our probe could be extended to higher dimensions: a spinning surface might have a role to play in computing entanglement entropy in higher dimensional field-theories with gravitational anomalies.   Of course, gravitational anomalies only occur in $d=4k+2$ dimensions $(k\in \mathbb{Z})$, so the only interesting case is $d=6$ QFTs. More interesting is the case of mixed gauge-gravitational anomalies,\footnote{We thank A. Wall for discussions on this point.} which occur in all even dimensions $d>2$. (In $d=2$, the would-be anomaly polynomial, $\Tr F \Tr R$, vanishes.) The holographic description of such anomalies is well known \cite{Witten:1998qj}, and their contribution to entanglement entropy can be calculated using our methods. More speculatively still, the geometrical character of expressions such as \eqref{finalS} suggests that there may be an interplay between entanglement and anomalies that extends beyond holography.

\item In 3d gravity, one can actually compute the bulk path integral $Z_n$ with replica boundary conditions at infinity in a semiclassical approximation, for any number of intervals. This was done for Einstein gravity in \cite{Faulkner:2013yia}, and it would be nice to explicitly generalize this to TMG. For locally AdS$_3$ manifolds at least, computing $Z_n$ should be tractable \cite{Faulkner:2013yia, Perlmutter:2013paa}. This should yield the R\'enyi entropies discussed in Section 2 for holographic CFTs with $c_L\neq c_R$. Doing so would provide an orthogonal and complementary method to those employed herein, and comes equipped with a prescription for computing bulk loop corrections to the classical result \cite{Chen:2014kja}.

\item The Mathisson-Papapetrou-Dixon equations admit, in principle, multiple solutions for the trajectory of the particle even for empty AdS$_3$. It is not clear if these additional saddles imply ambiguities like those found in \cite{Erdmenger:2014tba}. It would be interesting to construct these solutions explicitly and understand their role (if any). However, in the Chern-Simons formulation there is no such ambiguity: for flat connections, the equations of motion are first order in the momenta of the particle and the action is primarily
 sensitive to boundary conditions of the probe. This is carried out explicitly in Appendix \ref{app:CS}. For this reason we expect that non-geodesic solutions to \eqref{eq:p33d} either don't satisfy our boundary conditions or are subdominant when minimizing \eqref{repEE}.

\end{enumerate}

We hope to return to some of these issues in the future. Recently holography has permitted a refined understanding of the physical consequences of entanglement and anomalies both: we hope that it may have something to say also about the interplay of these two fundamental ideas in quantum field theory, and that our work may be viewed as a small step in that direction.

\begin{acknowledgements} 

It is a pleasure to acknowledge helpful discussions with M. Ammon, J. Armas, T. Azeyanagi, J. Camps, X. Dong, T. Hartman, M. Headrick, K. Jensen, M. Kulaxizi, A. Lawrence,  R. Loganayagam, G.-S. Ng and A. Wall. In particular, we are very grateful to M. Ammon for his early contribution to this collaboration. We are also grateful to the participants of the Solvay Workshop on ``Holography for Black Holes and Cosmology''. Finally, we wish to thank J. Camps, K. Jensen and G.-S. Ng for helpful comments on a draft. S.D. is a Research Associate of the Fonds de la Recherche Scientifique F.R.S.-FNRS (Belgium). E.P. has received funding from the European Research Council under the European Union's Seventh Framework Programme (FP7/2007-2013), ERC Grant agreement STG 279943, ``Strongly Coupled Systems''. N.I. is supported in part by the NSF under Grant No. PHY11-25915 and by the DOE under Grant No. DE-FG02-91ER40618.
\end{acknowledgements}

\appendix

\section{Conventions}

Our conventions for the epsilon tensor  $\epsilon_{\mu\nu\lambda}$ in Lorentzian signature with coordinates $(t,x,z)$ are: 
\be
\epsilon_{txz}=\sqrt{-g}~,\quad \epsilon^{txz}=-{1\over \sqrt{-g}}~. 
\ee
 Some other useful properties of the tensor are
\bea
\epsilon_{\mu \nu\lambda}\epsilon^{\mu \alpha \beta}= -\delta^\alpha_\nu\delta^\beta_\lambda+\delta^\alpha_\lambda\delta^\beta_\nu~, \quad
\nabla \epsilon_{\mu\nu\lambda}=0 ~.
\eea
When analytically continuing from Euclidean time $\tau_E$ to Lorentzian time $t$ we use
\be
\tau_E = i t \ .
\ee
This means that if ever we need to compare a Euclidean epsilon tensor with a Lorentzian one, they are related as
\be
\epsilon^{\mu\nu\rho}\big|_{E} = i \epsilon^{\mu\nu\rho}|_{L} \ .
\ee
We define symmetric and anti-symmetric tensors with factors
\be
a_{[\mu\nu]}={1\over 2}(a_{\mu\nu}-a_{\nu\mu})~, \quad a_{(\mu\nu)}={1\over 2}(a_{\mu\nu}+a_{\nu\mu})~.
\ee
We define the Christoffel symbol and spin connection as
\be\label{eq:gg}
\Gamma^\alpha_{\mu\nu}={1\over 2} g^{\alpha\beta}\left[\partial_\mu g_{\nu\beta}+\partial_\nu g_{\mu\beta}-\partial_\beta g_{\mu\nu}\right]~,
\ee
\be
\omega^a_{~b\mu}=e^{a}_{~\nu}\left(\partial_\mu e^{~\nu}_{b}+\Gamma^{\nu}_{\mu\lambda} e^{~\lambda}_{b}\right)~.
\ee
The Riemann tensor is 
\be
R^\rho_{~\sigma \mu\nu}=\partial_\mu \Gamma^\rho_{\nu\sigma}-\partial_\nu \Gamma^\rho_{\mu\sigma} +\Gamma^\rho_{\mu\lambda}\Gamma^\lambda_{\nu\sigma}-\Gamma^\rho_{\nu\lambda}\Gamma^\lambda_{\mu\sigma}
\ee
Locally AdS$_3$ spaces satisfy
\be\label{eq:RR}
R_{\mu\nu\rho\sigma}=-{1\over \ell^2}(g_{\mu\rho}g_{\nu\sigma}-g_{\nu\rho}g_{\mu\sigma})
\ee
with $\ell$ the AdS radius.

The covariant derivative  along a curve $X^\mu(s)$ is
\be\label{def:cov}
\nabla V^\mu:= {d V^\mu\over d s}+\Gamma^{\mu}_{\lambda\rho}\dot X^\rho V^\lambda~,
\ee
which, in terms of the one-form spin connection $\omega$, is
\be\label{def:do}
\nabla V = {d V\over d s} + [\omega, V]~, \quad V=V^a J_a ~, \quad V_a= V_\mu e^{~\mu}_a~.
\ee
where $J_a\in sl(2,\RR)$ and 
\be
\omega= \omega_{~\mu}^a J_a\, \dot X^\mu  ~,\quad \omega_{~\mu}^a=-\epsilon^{abc}\omega_{bc\mu}~.
\ee
When we introduce a Lorentzian normal frame $(n, \tn)$ to a spacelike curve with tangent vector $v^{\mu}$, we pick the following handedness:
\be
\tn^{\mu} = \ep^{\mu\nu\rho} v_{\nu} n_{\rho},
\ee
where $n^2 = -1$ and $\tn^2 = v^2 = 1$. The spin tensor $s^{\mu\nu}$ is defined as
\be
s^{\mu\nu} = \fs \le(n^{\mu} \tn^{\nu} - n^{\nu} \tn^{\mu}\ri) = - \fs\,\ep^{\mu\nu\rho} v_{\rho} \ .
\ee

\section{Details of cone action} \label{app:coneac}
In this section we compute the gravitational Chern-Simons action 
\be
S_{\rm CS} := \int d^3x\sqrt{-g}\epsilon^{\lambda\mu\nu}\Gamma^\rho_{\lambda\sigma}\left(\partial_\mu \Gamma^\sigma_{\rho\nu}+{2\over 3}\Gamma^\sigma_{\mu\tau}\Gamma^\tau_{\nu\rho}\right)
\ee
on a regularized cone, given by the metric \eqref{conemetnice}, which we write as
\be
ds^2 = e^{\ep \phi(\sig)} \delta_{ab} d\sig^a d\sig^b + \le(g_{yy} + K_a \sig^a + \cdots \ri) dy^2 + e^{\ep \phi(\sig)} U_a(\sig,y) d\sig^a dy \ . \label{conemet1app}
\ee
Here the $\sig^a$ are flat Cartesian coordinates on the space transverse to the cone. We denote the total opening angle of the cone by $2\pi n$, where we parametrize small deviations near $n \sim 1$ with $\ep = (n-1)$ and we will compute the action to first order in $\ep$. $\phi(\sig)$ is a regulatory function that smoothens out the tip of the cone. 

There is an important subtlety in this parametrization that is relevant for TMG: we would like the information regarding the opening angle of the cone to be stored in the periodicity of the coordinates $\sig^a$, and not in the metric components. If we go to complex coordinates $z := \sig^1 + i \sig^2$, then we have the following identification pattern
\be
z \sim z e^{2\pi i n} \ . \label{zident}
\ee 
Operationally, this means that the regulator function $\phi(\sig)$ falls off exponentially outside a small core that can be taken to have size $a$. From the point of view of computing a holographic partition function the asymptotic periodicity of the coordinates $z$ ultimately determines the periodicity of the coordinate system describing the dual CFT. To compute the entropy we should compute the CFT partition function given an identification such as \eqref{zident} and then take a derivative with respect to $n$. In a diffeomorphism-invariant bulk theory we are allowed to change coordinates, storing the information regarding the opening angle instead in the metric components (for example, see \eqref{conemet1} in Appendix \ref{app:back}). In TMG this is not actually physically equivalent, and such a procedure may result in different answers.

We now discuss the explicit form of the regulator function. The coordinate $w := z^{\frac{1}{n}}$ has the usual periodicity $w \sim w e^{2\pi i}$, and so the metric $dw d\bar{w} \sim  (z\zb)^{\frac{1}{n}-1} dz d\zb$ is regular at the origin. We seek to interpolate between this metric and $dz d\zb$ far from the tip of the cone: a suitable 2d metric that does this is
\be
dz d\zb \le( \le((z \zb)^{\frac{1}{n}-1} - 1\ri)f(\sqrt{z \zb}) + 1\ri)
\ee
with $f(r) := \exp\le(-\frac{r}{a}\ri)$. To first order in $\ep$ this takes the form \eqref{conemet1app} with the identification 
\be
\phi(z,\zb) = -f\le(\sqrt{z \zb}\ri) \log\le(z\zb\ri) \ . \label{expphi}
\ee

We now directly compute the action. The piece which survives in the $a \to 0$ limit takes the form
\be
S_{\rm cone, CS} = \frac{\ep}{4}\int d^2 \sig dy \le((-\delta^{ab}\p_a \p_b \phi) \ep^{ab} \p_{a} U_b + (\p_a \phi) \delta^{ab}\p_b (\ep^{cd} \p_c U_d)\ri) \ .
\ee
In evaluating this we have neglected terms nonlinear in $U_a$: this is because there is no loss in generality in taking $U_a(\sig = 0) = 0$, and thus these terms vanish when evaluated at the tip of the cone. We now integrate by parts on the second term, to find
\be
S_{\rm cone, CS} = -\frac{\ep}{2}\int d^2 \sig dy \le((\delta^{ab}\p_a \p_b \phi) \ep^{ab} \p_{a} U_b \ri) \ .
\ee
This integration by parts is justified only if $\phi(\sig)$ is taken to have compact support, which means that the $\epsilon$ variation actually involves a variation of the periodicity of the coordinates as in \eqref{zident}. 

Using the explicit form of \eqref{expphi} we now evaluate the integral over the $\sig^a$ to find
\be
S_{\rm cone, CS} = - 2\pi \ep \int dy \le(\ep^{ab} \p_{a} U_b \ri)~.
\ee
This is the result quoted in \eqref{CSexp}. 

\section{Details of variation of probe action}\label{app:p3}

In this Appendix we vary the action of the spinning particle \eqref{eq:actp3} with respect to the position of the particle worldline $X^{\mu}(s)$ and demonstrate that the resulting equations of motion are equivalent to the Mathisson-Papapetrou-Dixon equations of motion for a spinning particle in general relativity. The action is
\be
S_{\rm probe}=S_{\rm geod}+ S_{\rm anom} + S_{\rm constraints}~,
\ee
 with
 \bea
S_{\rm geod}&=& {\mathfrak{m}}\int_C ds \, \sqrt{g_{\mu\nu}\dot X^\mu  \dot X^\nu}~, \cr
S_{\rm anom}&=& \mathfrak{s} \int_C ds\, \tilde{n} \cdot \nabla n ~,\cr
 S_{\rm constraints}&=& \int_C ds \left[\lambda_1 n \cdot \tilde{n} + \lambda_2 n \cdot \dot{X} +\lambda_3 \tilde{n} \cdot \dot{X} +\lambda_4 (n^2+1) +\lambda_5(\tilde{n}^2-1)\right]~.
 \eea
We will work work in a parametrization always where $s$ measures proper distance and so we have $v^{\mu} = \frac{d X^{\mu}}{ds}, v^2 = 1$. To simplify our results we will use the equations of motion from the variation of the action with respect to  $(n,\tilde{n})$, which were worked out to be:
\bea 
-\mathfrak{s}\nabla \tn^\mu +\lambda_1 \tn^{\mu}+\lambda_2 v^\mu +2\lambda_4 n^{\mu} & = 0 \label{neqapp1} \\
\mathfrak{s}\nabla n^{\mu}+\lambda_1 n^{\mu}+\lambda_3 v^\mu +2\lambda_5 \tn^{\mu} & = 0, \label{neqapp2}
\eea
 We vary first with respect {\it only} to the explicit dependence on the coordinates. The contribution from the geodesic is the usual one\footnote{We ignore total derivatives in this section.}
 \bea\label{eq:vg}
 \delta_X S_{\rm geod}&=& {\mathfrak{m}}\int_C ds \, \delta_X \le(\sqrt{g_{\mu\nu}\dot X^\mu  \dot X^\nu}\ri)\cr
  &=& - \mathfrak{m}\int_C ds \,(\nabla v_\sigma) \delta X^\sigma~.
 \eea
 
 For the terms involving the normal vectors we have
 \bea\label{eq:vn}
 \delta_X S_{\rm anom}&=& \mathfrak{s} \int_C ds\, \delta_X( g_{\mu\nu} \tilde{n}^\nu  \nabla n^\mu) \cr
 &=& \mathfrak{s} \int_C ds\,  \left[  (\partial_\sigma g_{\mu\nu} )\tn ^\nu n^\mu+ (\partial_\sigma \Gamma^\mu_{\alpha\beta}) n^\alpha \tn_{\mu} v^\beta -\partial_s( \Gamma^\mu_{\alpha\sigma} n^\alpha \tn_{\mu} )\right]\delta X^\sigma \cr
 &=&\mathfrak{s} \int_C ds\,  \left[(\partial_\sigma g_{\mu\nu}) \tn^\nu \nabla n^\mu + R^\mu_{~\alpha\sigma\beta} n^\alpha \tn_{\mu} v^\beta -\nabla(n^\alpha \tn_{\mu} ) \Gamma^\mu_{\alpha\sigma} \right]\delta X^\sigma 
 \eea
 The variations of the constraints with respect to $x^\mu(s)$ gives
\bea\label{eq:vc}
\delta_X S_{\rm constraints}&=& \int_C ds \, \delta_x\left( g_{\mu\nu} \left[\lambda_2 n ^\mu \dot{X}^\nu +\lambda_3 {\tn}^\mu \dot{X}^\nu +\lambda_4 n^\mu n^\nu +\lambda_5 \tn^\mu \tn^\nu\right]\right)\cr
&=& \int_C ds \,  \partial_\sigma g_{\mu\nu} \left[\lambda_2 n^\mu \dot{X}^\nu +\lambda_3 \tn^\mu \dot{X}^\nu +\lambda_4 n^\mu n^\nu +\lambda_5 \tn^\mu \tn^\nu\right]\delta X^\sigma \cr
&& -  \int_C ds \, \frac{d}{ds} \left(\lambda_2 n_{\sigma}  +\lambda_3 \tn_{\sigma}  \right)\delta X^\sigma  \cr
&=& \int_C ds \, \left[ \lambda_2  v^\nu \left(n^\mu\partial_\sigma g_{\mu\nu}-n_{\alpha}\Gamma^\alpha_{\sigma\nu}\right)  +\lambda_3  v^\nu \left(\tn^\mu\partial_\sigma g_{\mu\nu}-\tn_{\alpha}\Gamma^\alpha_{\sigma\nu}\right) \right]\delta X^\sigma \cr
&&  + \int_C ds \, \left[ \partial_\sigma g_{\mu\nu}(\lambda_4 n^\mu n^\nu +\lambda_5 \tn^\mu \tn^\nu)- \nabla \left(\lambda_2 n_{\sigma}  +\lambda_3 \tn_{\sigma}  \right)\right]\delta X^\sigma ~.
\eea
Here we have used the fact (following from \eqref{neqapp1} and \eqref{neqapp2}) that on-shell we have $\lam_1 = 0$.  
 
In \eqref{eq:vn} and \eqref{eq:vc} there are some non-tensorial terms which we may be justified in labeling `{\it unwanted}'. Let us now gather together these terms and show that they add up to zero:
 \bea
 {\rm `unwanted'}&=&  \mathfrak{s} \int_C ds\,  \left[(\partial_\sigma g_{\mu\nu}) \tn^\nu \nabla n^\mu -\nabla(n^\alpha \tn_{\mu} ) \Gamma^\mu_{\alpha\sigma} \right]\delta X^\sigma\cr
 &&+ \int_C ds \, \left[ \lambda_2  v^\nu \left(n^\mu\partial_\sigma g_{\mu\nu}-n_{\alpha}\Gamma^\alpha_{\sigma\nu}\right)  +\lambda_3  v^\nu \left(\tn^\mu\partial_\sigma g_{\mu\nu}-\tn_{\alpha}\Gamma^\alpha_{\sigma\nu}\right) \right]\delta X^\sigma \cr
&&  + \int_C ds \, \left[ \partial_\sigma g_{\mu\nu}(\lambda_4 n^\mu n^\nu +\lambda_5 \tn^\mu \tn^\nu)\right]\delta X^\sigma ~.
 \eea
  Next we use \eqref{neqapp1} and \eqref{neqapp2} to replace $\nabla n, \nabla \tn$ for linear combinations of $(n,\tn,v)$; the terms proportional to $\lambda_3$ cancel automatically and the remaining terms are 
    \bea
 {\rm `unwanted'}&=&  \int_C ds\, (\lambda_4 n^\mu n^\nu+\lambda_5 \tn^\mu \tn^\nu) (2 g_{\alpha\nu}  \Gamma^\alpha_{\mu\sigma}-\partial_\sigma g_{\mu\nu}) \delta X^\sigma\cr
 &&  +\int_C ds\, \lambda_2 v^\nu n^\mu\left(\partial_\sigma g_{\mu\nu}-g_{\alpha\mu}\Gamma^\alpha_{\sigma \nu}-g_{\alpha\nu}\Gamma^\alpha_{\sigma \mu}\right)\delta X^\sigma\cr
 &=&0~.
 \eea
To obtain the second equality we used the definition of the Levi-Civita connection in terms of the metric \eqref{eq:gg}. 

We thus have
\bea
\delta_X S_{\rm probe}&=&\delta_X S_{\rm geod}+\delta_X S_{\rm anom} +\delta_X S_{\rm constraints}\cr
&=&  \int_C ds \left[-m \nabla v_\sigma +\mathfrak{s} R^\mu_{~\alpha\sigma\beta} n^\alpha \tn_{\mu} v^\beta -\nabla \left(\lambda_2 n_{\sigma}  +\lambda_3 \tn_{\sigma}  \right) \right] \delta X^\sigma~.
\eea
Our final step is to write the terms involving normal vectors as functions of the spin tensor $s^{\mu\nu}$. As in \eqref{def:s} we define
\be
s^{\mu\nu}= \mathfrak{s} \left(n^\mu \tn^\nu -\tn^\mu n^\nu\right) ~.
\ee
Using \eqref{neqapp1} and \eqref{neqapp2} with the above definition, we find the following relations:
\bea\label{eq:vsv}
&&\lambda_2 n^\sigma  +\lambda_3 \tn^{\sigma}   = v_\mu \nabla s^{\sigma \mu}~,\cr
&&\mathfrak{s} R_{\nu\mu\sigma\beta}\, n^\mu \tn^{\nu} v^\beta= -{1\over 2}R_{\sigma\beta \mu \nu} s^{\mu\nu} v^\beta~,
\eea
and hence setting $\delta_X S_{\rm MPD}=0$ gives us the final equation
\be
\boxed{\nabla\left[\mathfrak{m} v^\sigma + v_\mu \nabla s^{\sigma\mu}\right]=-{1\over 2}v^\beta s^{\mu\nu}R^\sigma_{~\beta\mu\nu}}~,
\ee
which is the Mathisson-Papapetrou-Dixon equation. 

As a final remark, one may be concerned that to demonstrate the covariance of the equations of motion we had to use the on-shell equations of motion for $(n, \tn)$, as surely the {\it full} set of the equations of motion should be covariant even off-shell. In fact we have been a bit quick. In reality a variation of the particle path also induces a variation of the normal vectors $n^{\mu}, \tn$: they should be parallel transported to the new location by demanding $\delta X^{\nu} \nabla_{\nu} n = 0$, i.e.
\be\label{eq:dn}
\delta_X n^\mu = - \Gamma^\mu_{\alpha\beta}n^\alpha \delta X^\beta~.
\ee 
By including this variation we can find a set of equations of motion that is covariant off-shell and reduces to the MPD equations on-shell. However a quicker route to the MPD equations is to note that if the equations of motion \eqref{neqapp1} and \eqref{neqapp2} are already satisfied, then the action is already stationary with respect to {\it any} variation of $n$, and thus the variation \eqref{eq:dn} may be neglected. 

\section{Details of backreaction of spinning particle} \label{app:back}

In this section we spell out some details on the relation between the Mathisson-Papapetrou-Dixon equations and the construction of bulk solutions to topologically massive gravity with a conical defect. 

\vskip 0.1 in
{\bf 1. Singular terms in TMG equations of motion: boundary condition method}
\vskip 0.1 in

As discussed in detail in Section \ref{sec:coneyisland}, to compute entanglement entropy we are attempting to construct solutions to the Euclidean TMG equations of motion
\be
R_{\mu\nu} - \frac{1}{2} g_{\mu\nu} R - \frac{1}{\ell^2} g_{\mu\nu} + \frac{i}{\mu} C_{\mu\nu} \equiv \sG_{\mu\nu} = 0 \label{eucEOMapp}
\ee
that contain a conical defect along a curve $C$. If we pick coordinates so that $C$ lies along $(z,\zb) = 0$ then the most general metric near $C$ is of the form \cite{Dong:2013qoa, Camps:2013zua}:
\bea
   ds^2 = e^{2A}[dz d\zb + e^{2A} T(\zb dz - zd\zb)^2]  & + (g_{yy} + 2 K_z z + 2 K_{\zb} \zb + Q_{zz} z^2 + Q_{\zb \zb} \zb^2 + 2 e^{2A} Q_{z \zb} z \zb   ) dy^2 \nonumber\\
  &  + 2 i  e^{2A}(U + V_z z + V_{\zb} \zb) (\zb dz - z d\zb) dy + \cdots, \label{conemet1}
\eea
where all of the functions appearing here are allowed to depend on $y$, the direction perpendicular to the cone. $A(\rho)$ is a function which regulates the tip of the cone, i.e.
\be
A(\rho) \equiv -\frac{\epsilon}{2} \log (\rho^2 + a^2) \ . \label{Adef}
\ee

 We would now like to show that if we examine these equations of motion, they will generically contain divergences as $\rho \to 0$; these divergences cannot be canceled by matching on to the rest of the geometry and we should demand that their coefficients vanish. This constrains the set of $C$ that can support such a singularity. In the case of pure Einstein gravity, this procedure requires the vanishing of the traces of the extrinsic curvature and thus requires that $C$ be a minimal surface \cite{Lewkowycz:2013nqa}. In TMG, we will show that it instead requires that $C$ satisfy (the Euclidean analytic continuation of) the Mathisson-Papapetrou-Dixon equations \eqref{eq:VV}, which follow from the extremization of the entanglement functional \eqref{totSeuc}. 

First we understand the geometric properties of $C$ as encoded in the metric \eqref{conemet1}. The velocity vector $v^{\mu}$ along $C$ is $v \equiv \frac{\p_y}{\sqrt{g_{yy}}}$, evaluated at $\rho = 0$. From this we may construct the acceleration vector
\be
a^{\mu} \equiv v^{\rho} \nabla_{\rho} v^{\mu} =  \frac{1}{g_{yy}}\le(- K_{\zb} \p_{z} - K_{z} \p_{\zb}\ri) \label{acc}
\ee 
and its covariant derivative, sometimes rather unfairly called the ``jerk'':
\be
j^{\mu} \equiv v^{\rho} \nabla_{\rho} a^{\mu} = \frac{1}{g_{yy}^{3/2}}\le(\le(-\p_{y} K_{\zb} + 2 i U K_{\zb}\ri)\p_{z} +\le(-\p_y K_z - 2 i K_z U \ri)\p_{\zb} - \frac{K_{\zb} K_{z}}{g_{yy}}\p_y\ri) \label{jerk} .
\ee
These expressions require us to evaluate $A(0) = -\ep\log a \to 0$ where the last equality is valid as $\ep \to 0$. We may set $\ep \to 0$ here with no loss of information as the geometric properties of $C$ do not depend on the strength of the singularity. 

Next we note that in terms of the acceleration and the jerk the Euclidean continuation of the MPD equations \eqref{eq:VV} take the form 
\be
\fm a^{\sig} + i\fs \le(\ep^{\sig\mu\rho} v_{\mu} j_{\rho} + \ha v^{\beta} \ep^{\mu\nu\rho}R^{\sig}_{\phantom{\sig}\beta\mu\nu} v_{\rho}\ri) = 0
\ee
The factors of $i$ arise from the fact that the epsilon tensor used here is the Euclidean signature version and our previous definition of the spin tensor was in Lorentzian signature. Using \eqref{acc} and \eqref{jerk} we find these equations become 
\be
\le(-\fm\,\sqrt{g_{yy}}  K_{\zb} + i \fs \,\p_{y}K_{\zb}\ri) + 2\fs\, \le(K_{\zb} U - 3 g_{yy} V_{\zb}\ri) = 0 \label{MPDkz}
\ee
and the corresponding equation with $z \leftrightarrow \zb$. Note if the spin is taken to $0$ this simply states that $K_{z} = K_{\zb} = 0$, i.e. the geodesic equation. 

Next, we simply directly evaluate the left-hand side of the equations of motion \eqref{eucEOMapp} using the cone ansatz. The $zz$ and $\zb\zb$ side of the equations of motion contain $1/\rho$ divergences, e.g.
\be
\sG_{\zb\zb} \sim -\frac{\ep}{2 \rho} \frac{ e^{i\tau}}{ g_{yy}^{\frac{3}{2}}}\le(K_{\zb} \sqrt{g_{yy}} - i \frac{1}{\mu} \p_y K_{\zb} - \frac{2}{\mu}\le(K_{\zb} U - 3 g_{yy} V_{\zb}\ri)\ri) + \sO(\ep^2)
\ee
This divergence will vanish precisely when the MPD equation \eqref{MPDkz} is satisfied, provided the ratio of the mass to the spin of the probe satisfies the relation
\be
\frac{\fs}{\fm} = \frac{1}{\mu} \ .
\ee
Thus we conclude that if the curve $C$ satisfies the MPD equations, then the singular part of the bulk equations of motion vanish and a geometry can be constructed with a conical defect along $C$. 

\vskip 0.1 in
{\bf 2. Matching delta functions}
\vskip 0.1 in

Here we consider a different route. Consider sourcing the TMG action with a spinning particle by studying the combined action
\be
S = S_{\rm TMG}[g] + \ep S_{\rm probe}[g,X,n]
\ee
This results in the Euclidean equations of motion
\be
\frac{1}{8\pi G_3}\le(R^{\mu\nu} - \ha g^{\mu\nu} R - \frac{1}{\ell^2} g^{\mu\nu} + \frac{i}{\mu} C^{\mu\nu}\ri) + \ep T^{\mu\nu}_{\rm probe} = 0 \label{sourcedTMGeom},
\ee
where we have defined the stress tensor of the probe by
\be
\delta S_{\rm probe} =  \int d^3x \sqrt{g} \le( \ha T^{\mu\nu}_{\rm probe} \delta g_{\mu\nu} \ri) \ .
\ee
In this section we will show that the resulting spacetime sourced by the spinning probe is the desired conical defect studied extensively above. The resulting equations are then consistent only if we minimize the action of the probe over the dynamical variables $X^{\mu}$ and $n^i$. To understand this consider taking the divergence of  \eqref{sourcedTMGeom}; the geometric terms vanish as an identity, and so we find
\be
\nabla_{\mu} T^{\mu\nu}_{\rm probe} = 0
\ee
This conservation law is equivalent to the MPD equations\footnote{While we have checked it explicitly in this case, the fact that the stress-tensor of a brane is conserved if the action of the brane is extremized with respect to all dynamical variables follows from bulk diffeomorphism invariance. In this context it helps us to understand why we must vary with respect to the normal vectors as well as with respect to the particle trajectory.}. Thus we need only verify that the backreaction of the probe creates the desired singularity in the spacetime.  

We will do this in two steps: first we compute the Euclidean stress tensor arising from the probe. In Euclidean signature $S_{\rm probe}$ takes the form
\be
S_{\rm probe}=S_{\rm geod}+ S_{\rm anom} + S_{\rm constraints}~,
\ee
 with
 \bea
 S_{\rm geod}&=& {\mathfrak{m}}\int_C ds \, \sqrt{g_{\mu\nu}\dot X^\mu  \dot X^\nu}~, \\
S_{\rm anom}&=& i \mathfrak{s} \int_C ds\, n_2 \cdot \nabla n_1 ~,\\
 S_{\rm constraints}&=& i \int_C ds \left[\lambda_1 n_1 \cdot n_2 + \lambda_2 n_1 \cdot \dot{X} +\lambda_3 n_2 \cdot \dot{X} +\lambda_4 (n_1^2-1) +\lambda_5(n_2^2-1)\right].
 \eea
In this Appendix some of the $\lam_i$ have been redefined with factors of $i$ relative to other sections of this paper. This metric variation can be divided into two parts, as $S_{\rm probe}$ depends on the metric both explicitly and through the definition of the Christoffel symbol $\Gamma_{\mu\nu}^\alpha$. Thus consider first variations that treat them separately
\be\label{eq:svar}
{\delta S_{\rm probe} }= \int d^3y \sqrt{g} \left({1\over 2}t^{\mu\nu}\delta g_{\mu\nu} + L^{\mu\nu}_{\phantom{ab}\alpha}\delta \Gamma^{\alpha}_{\mu\nu}\right)~.
\ee
Taking into account the relation between $\Gamma^\alpha_{\mu\nu}$ and the metric
\be
\delta \Gamma^\alpha_{\mu\nu}={1\over 2}\left(g^{\alpha \beta}\nabla_\mu \delta g_{\nu\beta}+g^{\alpha \beta}\nabla_\nu \delta g_{\mu\beta}-\nabla^\alpha\delta g_{\mu\nu}\right)~,
\ee
we find that the full stress tensor is
\bea\label{tmpd}
T_{\rm probe}^{\mu\nu}= t^{\mu\nu}- \nabla_\alpha\left(L^{\nu[\alpha \mu]}+L^{\mu[\alpha\nu]}+L^{\alpha(\mu\nu)}\right)~.
\eea
We now need to actually construct this variation. For convenience we define the notation
\be
\int {\bf ds}\equiv \int_C ds\, {\delta^{(3)}(y-X(s))\over \sqrt{g}},
\ee
and begin by constructing the tensor $t_{\mu\nu}$ as defined in equation \eqref{eq:svar}:
\bea\label{eq:tm}
t^{\mu\nu}&=&\int {\bf ds}\left[\mathfrak{m} v^{\mu} v^{\nu} + i \le(2 \mathfrak{s}n_2^{(\mu}\nabla n_1^{\nu)} + 2\lambda_2 n_{1}^{(\mu}v^{\nu)} +2\lambda_3 n_{2}^{(\mu}v^{\nu)}+2\lambda_4 n_1^\mu n_1^\nu +2\lambda_5 n_2^\mu n_2^\nu\ri)\ri] \cr
&=&  \int {\bf ds}\left[\mathfrak{m} v^{\mu} v^\nu+ i \le(\mathfrak{s}\nabla(n_1^{(\mu}n_2^{\nu)})+ \lambda_2 n_{1}^{(\mu}v^{\nu)} + \lambda_3 n_{2}^{(\mu}v^{\nu)}\ri)\right]\cr
&=& \int {\bf ds}\left[\mathfrak{m} v^\mu v^\nu +i \le(\mathfrak{s}\nabla(n_1^{(\mu}n_2^{\nu)})+v_\alpha v^{(\mu}\nabla s^{\nu)\alpha}\ri)\right] 
\eea
From the first to the second line we used the Euclidean analogs of \eqref{neqapp1} and \eqref{neqapp2} arising from the variation of the action with respect to $n_i$; from the second to the third line we used \eqref{eq:vsv} but with the spin tensor defined as
\be
s^{\mu\nu} = \fs \le(n_1^{\mu} n_2^{\nu} - n_2^{\mu} n_1^{\nu}\ri) \ .
\ee

The contribution to the variation of  $S_{\rm probe}$ coming directly from explicit dependence on the Christoffel symbol is 
\be
L^{\mu\nu}_{\phantom{ab}\alpha}=i\mathfrak{s} \int {\bf ds}\,  v^{\mu} n^{\nu}_1 n_{2\alpha}~.
\ee
In particular we find that 
\be
L^{\mu[\nu\alpha]}={i\over 2}\int {\bf ds} \, v^\mu s^{\nu\alpha}~.
\ee
We want to evaluate the contribution of $L^{\mu\nu}_{\phantom{ab}\alpha}$ to the total stress tensor as given by \eqref{tmpd}. To start, consider the term
\be\label{eq:DL}
\nabla_\alpha L^{\alpha \mu\nu} =\partial_\alpha L^{\alpha\mu\nu} + \Gamma^\alpha_{\alpha\beta} L^{\beta\mu\nu} +
 \Gamma^\mu_{\alpha\beta} L^{\alpha\beta\nu} +  \Gamma^\nu_{\alpha\beta} L^{\alpha\mu\beta}~.\ee
The first term is
\be\label{eq:LL}
\partial_\alpha L^{\alpha\mu\nu} =i\mathfrak{s} \int {ds}\,  v^{\alpha} n^{\mu}_1 n_{2}^{\nu} {\partial\over \partial y^\alpha}\left({\delta^{(3)}(y-x(s))\over \sqrt{g}}\right)~.
\ee
Using the following identities
\bea
{\partial\over \partial y^\alpha}\left( {1\over \sqrt{g}}\right)=- { \Gamma^\beta_{\alpha\beta}\over \sqrt{g}}~,\quad {\partial\over \partial y^\alpha}\left({\delta^{(3)}(y-x(s))}\right)= - {\partial\over \partial x^\alpha}\left({\delta^{(3)}(y-x(s))}\right)~,
\eea
equation \eqref{eq:LL} simplifies to
\bea
\partial_\alpha L^{\alpha\mu\nu} &=& - \Gamma^\beta_{\alpha\beta}L^{\alpha\mu\nu} - i\mathfrak{s} \int {ds}\,   n^{\mu}_1 n_{2}^{\nu} {\partial\over \partial s}\left(\delta^{(3)}(y-x(s))\right) {1\over \sqrt{g}}\cr
&=& - \Gamma^\beta_{\alpha\beta}L^{\alpha\mu\nu} + i\mathfrak{s} \int {\bf ds}\,  \frac{d}{ds}( n^{\mu}_1 n_{2}^{\nu}) ~.
\eea
Combining this result with \eqref{eq:DL} gives
\bea
\nabla_\alpha L^{\alpha (\mu\nu)} = i\mathfrak{s} \int {\bf ds}\,  \nabla( n^{(\mu}_1 n_{2}^{\nu)}) ~.
\eea
After the dust settles, we find
\be\label{eq:DDL}
 \nabla_\alpha\left(L^{\nu[\alpha \mu]}+L^{\mu[\alpha\nu]}+L^{\alpha(\mu\nu)}\right) =  i\mathfrak{s} \int {\bf ds}\,  \nabla( n^{(\mu}_1 n_{2}^{\nu)}) -  i\nabla_\alpha \left( \int {\bf ds}  \, v^{(\mu}s^{\nu)\alpha} \right)~.
\ee
Adding \eqref{eq:tm} and \eqref{eq:DDL} gives us for the total stress tensor:
\bea\label{eq:ttt}
\boxed{T^{\mu\nu}_{\rm probe}= \int {\bf ds}\left[ \mathfrak{m}\dot v^\mu \dot v^\nu + i v_\alpha v^{(\mu}\nabla s^{\nu)\alpha}\right] + i \nabla_\alpha \left( \int {\bf ds}  \, v^{(\mu}s^{\nu)\alpha} \right)~ \ .}
\eea 
The conservation of this stress tensor is equivalent to the MPD equations, as one expects on general grounds. Note that the normal vectors themselves have vanished, and the full answer can be written in terms of the spin tensor. 

We would now like to show that the backreaction of this stress tensor on the geometry (as dictated by the TMG equations) creates a conical singularity:
\be
\frac{1}{8\pi G_3}\le(R^{\mu\nu} - \ha g^{\mu\nu} R - \frac{1}{\ell^2} g^{\mu\nu} + \frac{i}{\mu} C^{\mu\nu}\ri) + \ep T^{\mu\nu}_{\rm probe} = 0 \label{sourcedTMGeom}
\ee
As it is well known that the part proportional to $\fm$ sources a delta function in the Einstein tensor, we must show that the new spin-induced delta function sources the corresponding delta function structure in the Cotton tensor. To that end we consider the regulated cone ansatz \eqref{conemet1}, expanded only to first order in $(z,\zb)$:
\be
ds^2 = e^{2A(z,\zb)} dz d\zb + (g_{yy} + K_{z}z + K_{\zb} \zb) dy^2 \label{simpans}
\ee
In this calculation we will construct local equations of motion evaluated at a fixed value of $y$, and we can use the gauge freedom \eqref{gaugeU} to set $U$ to 0 at that point. Computing its Cotton tensor we find that the singular terms take the form
\be
C^{zy} = \frac{4}{g_{yy}^{\frac{3}{2}}}\le(K_{\zb} \p \overline{\p} A + g_{yy} \p \overline{\p}^2 A\ri) + \cdots \qquad C^{\zb y} = -(C^{zy})^{*}  
\ee  
where we have neglected terms that are regular at the tip of the cone or are higher order in $\ep$. 

Furthermore, constructing the velocity $v$ and the spin tensor following from \eqref{simpans} and evaluating the stress tensor \eqref{eq:ttt}, we find after considerable algebra
\be
\ep T^{zy}(x)_{\rm probe} = - \frac{2 i \fs }{\pi g_{yy}^{\frac{3}{2}}}\le(K_{\zb} \p \overline{\p} A + g_{yy} \p \overline{\p}^2 A\ri) \qquad T^{\zb y}_{\rm probe} = (T^{zy}_{\rm probe})^{*}   
\ee
Here for ease of comparison with the Cotton tensor we have chosen to represent the two-dimensional delta function using the function $A(z,\zb)$ defined in \eqref{Adef}, i.e. we have
\be
\lim_{a \to 0} \p \overline{\p} A = -\ep \pi \delta^{(2)}(z,\zb)
\ee
Comparing these with the equations of motion \eqref{sourcedTMGeom} we see that the delta function structure agrees provided that we have
\be
\fs = \frac{1}{4 \mu G_3},
\ee
in agreement with the previous results. 

It is interesting to note that a spinning particle in TMG creates a pure (i.e. non-spinning) conical defect, provided that its spin is tuned to the Chern-Simons coupling appropriately. Conversely, a spinless particle in TMG actually creates a {\it spinning} conical defect geometry \cite{Deser:1989ri}.

\section{Chern Simons formulation of TMG}\label{app:CS}

In this appendix we will construct the effective action for the spinning probe in Chern-Simons language using the techniques developed in \cite{Ammon:2013hba}.\footnote{See also \cite{deBoer:2013vca} for an different (but compatible) approach to this subject.} In doing so we will show explicitly how the Chern-Simons and gravitational prescriptions are equivalent. This will give a different perspective on how to construct a spinning particle and, for some readers, may clarify certain properties of the  metric-like construction. 

We start by writing the TMG as a Chern-Simons theory with a constraint \cite{Deser:1991qk, Carlip:2008qh}. Define\footnote{Our conventions for $sl(2,\RR)$ are $[J_a,J_b]= \epsilon_{abc}J^c$ and in the fundamental representation
\be
J_0= \left[\begin{array}{cc}1/2 & 0 \\ 0& -1/2\end{array}\right]~,\quad J_{+}= \left[\begin{array}{cc}0& 0 \\ -1& 0\end{array}\right]~,\quad J_{-}= \left[\begin{array}{cc}0 & 1 \\ 0& 0\end{array}\right]~.
\ee
}
\be\label{eq:ALR}
A_L=\left(\omega^a+{1\over \ell}e^a\right) J_a~,\quad  A_R=\left(\omega^a-{1\over \ell}e^a\right) \bar J_a~,
\ee
where $A_{L,R}\in sl(2,\RR)_{L,R}$. The TMG action \eqref{eq:atmg} can be rewritten as \cite{Chen:2011yx}
\be\label{tmgact}
S_{\rm TMG}=(1-{1\over \mu\ell})S_{\rm CS}[A_L]-(1+{1\over \mu\ell})S_{\rm CS}[ A_R]-{k\over 4\pi \mu\ell}\int {\rm Tr}(\beta\wedge (F_L- F_R))~,
\ee
with $F=dA +A\wedge A$, and the Chern-Simons action is
\be\label{eq:csact}
S_{\rm CS}[A]={k\over 4\pi }\int {\rm Tr}\left(A\wedge d A+{2\over 3}A\wedge A\wedge A\right)~.
\ee
The Chern-Simons level $k$ is related to the gravitational Planck length and AdS$_3$ radius via
\be
k={\ell\over 4G_3} ~.
\ee
It is convenient to define left and right levels,
\be\label{levels}
k_L ={c_L\over 6}= k\left(1-{1\over \mu\ell}\right)~, \quad k_R = {c_R\over 6}=k\left(1+{1\over \mu\ell}\right)~,
\ee
where $(c_L,c_R)$ are the Brown-Henneaux central charges \eqref{eq:clcr}.

In this notation, $\beta$ is a one-form that enforces the ``torsion free'' condition $F_L=F_R$. The addition of this constraint is not just a matter of aesthetics: the Lagrange multiplier $\beta$ is the massive degree of freedom that characterizes TMG. Without it, the theory remains a topological, parity-odd  Chern-Simons theory of gravity with no local degrees of freedom \cite{Witten:1988hc,Townsend:2013ela}.\footnote{There exist proposals for theories of topologically massive higher spin gravity, see e.g. \cite{Chen:2011yx,Bagchi:2011vr} . These furnish a dual description of the currents in holographic CFTs with $c_L\neq c_R$ and an extended conformal symmetry. Higher spin TMG actions are known in closed form only in first-order formulation: namely, they are identical to the first-order action \eqref{tmgact} of ordinary TMG, only with fields now valued in a higher spin algebra ${\cal G}\supset sl(2,R)$. The analysis in this appendix in principle extends in a straightforward manner to compute holographic entanglement entropy for these theories.}

\vskip .1 in
{\bf 1. Wilson lines}\vskip .1 in

Following the work of \cite{Witten:1989sx,Carlip:1989nz,deSousaGerbert:1990yp},  in \cite{Ammon:2013hba} it was argued that the Wilson line 
\be\label{eq:WRC}
W_{\cal R}(C)={\rm tr}_{\cal R} \left( {\cal P}\exp \int_C {\cal A}\right)~.
\ee
correctly captures the dynamics of a massive particle in AdS$_3$ when ${\cal A} \in so(2,2)\sim sl(2,\RR)_L\times sl(2,\RR)_R$, with an appropriate choice of representation  $\cal R$ for the gauge group $G=SO(2,2)$.

Our goal is to construct a massive and spinning probe in the Chern-Simons language and show that it captures the same physics as the MPD equations. We specialize to flat connections, i.e. locally AdS$_3$ geometries. For present purposes we highlight the following properties of \eqref{eq:WRC}:
\begin{enumerate}
\item We need a representation that carries the data of a {\it massive and spinning} particle in AdS$_3$. The natural unitary representations for the probe are {\it infinite-dimensional}. In particular the highest-weight representation of $\slt$, defined in the  via a highest-weight state $|h,\bar h\rangle$ satisfies
\bea\label{eq:lwrep}
J_1 |h,\bar h \rangle = 0~, \quad J_0 |h,\bar h \rangle = h | h,\bar h \rangle~,\quad
\bar J_1 |h,\bar h \rangle = 0~, \quad \bar J_0 |h,\bar h \rangle = \bar h | h,\bar h \rangle~,
\eea
Here ${J_{0,\pm1}}\in sl(2,\RR)_L$ and ${\bar J_{0,\pm1}}\in sl(2,\RR)_R$.  There is an infinite tower of descendants created by $(J_{-1})^n (\bar J_{-1})^{\bar n}$.  In this notation, the mass and spin of the probe are
\be\label{eq:cms}
\ell\mathfrak{m}= h+\bar h~,\quad \mathfrak{s}=\bar h-h~.
\ee
The construction in \cite{Ammon:2013hba} took $ \mathfrak{s}=0$. One of the purposes of this appendix is to show how to account for non-trivial spin in the representation.  
\item One can interpret $\cal R$ as  the Hilbert space of an auxiliary quantum mechanical system that lives on the Wilson line. This auxiliary system can be constructed as a path integral over some fields $U$ which have a global symmetry group $G$: we will pick the dynamics of $U$ so that upon quantization the Hilbert space of the system will be precisely the desired representation $\sR$.
More concretely, we replace the trace over $\sR$ by a path integral,
\be\label{aa:pathint}
W_{\cal R}(C)= \int {\cal D}U \exp[-S(U; {\cal A})_C] ~,
\ee
where the action $S(U; {\cal A})_C$ is design that  $G$ is a local symmetry of the probe $U$ along the curve $C$.
\end{enumerate}
For more details on this construction, we refer the reader to \cite{Ammon:2013hba}. 

To construct a Wilson line in the representation \eqref{eq:lwrep} it is useful to treat the connections $A_L$ and $A_R$ separately, i.e.
\be
W_{\sR}(C)= W_{L}(C) W_R(C) 
\ee
where
\be
W_{L}(C)=  {\rm tr}_{\cal R} \left( {\cal P}\exp \int_C {A}_L\right) =\int {\cal D}U_L \exp[-S_L(U_L; {A}_L)_C]
\ee
and an analogous definition for $W_R(C)$. The effective actions for each Wilson line are as follows.
The action for our ``left movers''  in $W_L(C)$ is\footnote{In what follows $A_{L,R}$ are understood to be the pullback of the connections along  $C $, i.e.   $A= A_{\mu}  {\dot x^\mu} $.}
\bea\label{eq:actL}
S_L= \int_C ds\, {\rm Tr}\left(P_L D_L U_L \, U_L^{-1}\right) + \lambda_L( {\rm Tr}\left(P_L^2\right)-c_{2})~,\quad D_L U_L = \partial_s U_L + A_L U_L~, 
\eea
which is invariant under
\be
U_L\to L(s) U_L~,\quad P_L\to L(s)P_L L^{-1}(s)~, \quad A_L\to L(s)(A_L+d)L^{-1}(s)  ~,
\ee
with $L(s)\in SL(2,\RR)$. The action for the ``right movers'' in $W_R(C)$ is
\be\label{eq:actR}
S_R= \int_C ds \,{\rm Tr}\left(P_R U_R^{-1} D_RU_R \right) + \lambda_R( {\rm Tr}\left(P_R^2\right)-\bar c_{2})~,\quad D_R U_R = \partial_s U_R - U_R A_R ~,
\ee
and the local symmetries are given by
\be
U_R\to U_R R(s)~,\quad P_R\to R^{-1}(s)P_R R(s)  \quad A_R\to R^{-1}(s)(A_R+d)R(s)  ~, 
\ee
where $R(s)\in SL(2,\RR)$. Here $P_R$ and $P_L$ are elements of the algebra and correspond to the conjugate momenta to $U_R$ and $U_L$, respectively. The information about the representation $\sR$ is encoded in the quadratic Casimirs via 
\be\label{eq:cbc}
c_{2}=2h(h-1)~,\quad \bar c_{2}=2\bar h (\bar h-1)~.
\ee
And for $P=P_aJ^a$, we have 
\be
\Tr (P^2)=2P_0^2-(P_{-1}P_1+P_1P_{-1})~. 
\ee

From the actions \eqref{eq:actL} and \eqref{eq:actR}, the equations of motion are
\bea\label{eq:eom1}
U_R^{-1}D_R U_R + 2\lambda_R P_R=0~,\quad
\partial_s P_R +[A_R, P_R]=0\\
D_L U_L U^{-1}_L + 2\lambda_L P_L=0~,\quad
\partial_s P_L +[A_L, P_L]=0 \label{eq:eom2a}
\eea
 Integrating out $P_L$ and $P_R$ gives the following equations for $U_L$ and $U_R$
 \bea\label{eq:L1}
 {d\over ds}\left(U^{-1}_R \dot U_R -A_R\right) - [U^{-1}_R \dot U_R, A_R]=0~,\\ 
 {d\over ds}\left( \dot U_LU^{-1}_L +A_L\right) -[ \dot U_LU^{-1}_L,A_L]=0~.\label{eq:R1}
 \eea

\vskip .2 in {\it a. Wilson lines and MPD equations} \vskip .2 in

To see that the dynamics of $U_{L,R}$ are equivalent to those of the MPD particle, it is convenient to write the MPD equations in  a way that makes explicit the $SL(2,\RR)$ structure. We start by rewriting the MPD equations in terms of the momenta of the particle,
\bea\label{eq:VV1}
\nabla p^\mu={1\over 2}v^\nu s^{\rho\sigma}R^\mu_{~\nu\rho\sigma}~,\\
\nabla s^{\alpha\beta}+v^\alpha p^\beta -v^\beta p^\alpha=0  \label{eq:ss}~,
\eea
where the canonical momentum is given by $p^\mu=m v^\mu +v_\rho \nabla s^{\mu\rho}$. 
Next, define
\be
s^c:={1\over 2}\epsilon^{abc}s_{ab}~, \quad S:= J_c s^c~,
\ee
where $J_a\in SL(2,R)$ and $s^{ab}=e^a_\mu e^b_\nu s^{\mu\nu}$. Then contracting  \eqref{eq:ss} with $\epsilon_{abc}J^c$ we get 
\be\label{eq:s1}
\nabla S + \epsilon_{abc} J^c v^a p^b =0~,
\ee
and using $[J_a,J_b]=\ep_{abc}J^c$ gives
\bea
\nabla S + [e, P] =0~,
\eea
where we introduced the notation
\be
P:= J_a p^a ~,\quad  e:= J_a v^a = J_a e^a_\mu   \dot x^\mu~.
\ee

Next, consider \eqref{eq:VV1} for spacetimes that are locally AdS$_3$. We find
\be\label{eq:p1}
\nabla p^\mu=-{1\over 2}v^\nu s^{\rho\sigma}R^\mu_{~\nu\rho\sigma}=v^\nu s^{\mu}_{~\nu} ~,
\ee
where we used \eqref{eq:RR}. An equivalent rewriting of \eqref{eq:p1} using tangent space indices is 
\be
\nabla p^a+v^b \epsilon^{a}_{~ b c}s^c =0   ~.
\ee
Multiplying by $J_a$ we get
\be
\nabla P + [e,S]=0~.
\ee

Summarizing, the MPD equations \eqref{eq:VV1} and \eqref{eq:ss} are equivalent to
\be\label{xx}
\nabla P + [e,S]=0~,\quad \nabla S+ [e,P]=0~.
\ee
Emphasis should be placed on the fact that this is only valid for spacetimes that are locally AdS$_3$. 

With this notation it is straight forward to compare with the Chern-Simons formulation. The equations of motion for the left and right moving momenta are given by \eqref{eq:eom1} and \eqref{eq:eom2a}. In terms of covariant derivatives, as defined by \eqref{def:do}, we can write these equations as
\be \label{eq:cprl}
\nabla P_R -[e,P_R]=0 ~, \quad \nabla P_L +[e,P_L]=0
\ee
where we used \eqref{eq:ALR}. %
Comparing \eqref{eq:cprl} with \eqref{xx} we identify\footnote{Note that this identification is not unique; the normalizations and relative signs are not fixed. This is simply because  we have not specified how the Chern-Simons variables map to spacetime variables.}
\be
P= P_R+P_L ~,\quad S= P_R-P_L~.
\ee
This proves that the conservation equations of the left and right moving momenta associated to the Wilson lines are equivalent to the MPD equations, provided the connections $A_{L,R}$ are flat.

\vskip .2 in {\it b. Backreaction of Wilson Lines}\vskip .2 in
To fix the Casimirs relevant for computing entanglement entropy, we consider the backreaction of our probe on the gravity background. We switch to Euclidean signature and consider the action 
\be\label{probeback}
S = i\left( 1-\frac{1}{\mu \ell}\right) S_{CS}[A_L] - i\left( 1+\frac{1}{\mu \ell}\right)S_{CS}[A_R] + S_L(U_L;A_L)_C +S_R(U_R;A_R)_C ~,
\ee
where $S_{CS}$ is given by \eqref{eq:csact}, and 
the probe is described by the actions $S_{L,R}$ in \eqref{eq:actL} and \eqref{eq:actR}. If we vary the action \eqref{probeback} with respect to the gauge fields $A_L$ and $A_R$  we obtain
\bea
\frac{k}{2\pi}\left( 1-\frac{1}{\mu \ell}\right) F_{L,\mu\nu}(x) &=&  i \int ds \frac{dx^{\rho}}{ds}\ep_{\mu\nu\rho}\delta^{(3)}(x - x(s)) P_L ~,\cr
 \frac{k}{2\pi}\left( 1+\frac{1}{\mu \ell}\right) {F}_{R,\mu\nu}(x)& =& i \int ds \frac{dx^{\rho}}{ds}\ep_{\mu\nu\rho}\delta^{(3)}(x - x(s)) P_R~. \label{Fsource}
\eea

In Euclidean signature it is convenient to introduce complex coordinates $z$ and $\bar{z}$ for the field theory coordinates $t$ and $\phi$ by 
\be\label{zcoord}
z = \phi + i t_E \, , \qquad\quad \bar{z} = \phi - i t_E \, . 
\ee
In this Appendix we use $\rho$ as the Fefferman-Graham radial cordinate. Our probe should be parameterized (at least close to the boundary) by $\rho(s) = s$. Thus $P_L(s)$ and $P_R(s)$ are independent of $s$  which have to satisfy
\be
{\rm Tr}(P_L^2(s)) = c_{2} \, , \qquad {\rm Tr}(P_R^2(s)) = \bar{c}_{2} \, , 
\ee
which is, for example, achieved by 
\be
P_L(s) = \sqrt{2 c_{2}} J_0 \, , \qquad P_R(s) = \sqrt{2 \bar{c}_{2}} J_0 \, .
\ee

In order to solve \eqref{Fsource} we introduce $a_{L,\rm source}$ and $a_{R,\rm source}$ by 
\bea
A_L & =L a_{L,\rm source}L^{-1}+LdL^{-1} ~, \quad L = e^{-\rho J_0} e^{-J_{1} z}~,\cr
A_R & =R^{-1} a_{R,\rm source}R+R^{-1}dR~, \quad R = e^{-J_{-1} \bar z} e^{-\rho J_0}~,
\eea
where the gauge functions ($L, R$) generate the asymptotics, whereas the coupling to the source is taken into account by
\bea\label{Fsourcesol}
a_{L, \rm source} &=&  \sqrt{\frac{c_{2}}{2}}\frac{1}{k \left( 1-\frac{1}{\mu \ell}\right)}\le(\frac{dz}{z} - \frac{d\bar{z}}{\bar{z}}\ri) J_0~ \, , \cr
a_{R, \rm source} &=&  \sqrt{\frac{\bar{c}_{2}}{2}}\frac{1}{k \left( 1+\frac{1}{\mu \ell}\right)}\le(\frac{dz}{z} - \frac{d\bar{z}}{\bar{z}}\ri) J_0~
\eea
This is indeed a solution to the equations \eqref{Fsource} because 
\be
\partial_z \left( \frac{1}{\bar{z}} \right) = \partial_{\bar z} \left( \frac{1}{z} \right) = \pi \delta^{(2)}(z,\bar{z}) = \pi \delta(t_E) \delta(\phi).
\ee
Moreover, the antisymmetric epsilon tensor in the coordinates $(\rho, z, \bar{z})$ is normalised such that 
\be
\epsilon_{\rho z \bar{z}} =\frac{i}{2} \, .
\ee
Let us now calculate the metric associated to the solution. It is convenient to write
\be\label{identc2andk}
\sqrt{2 c_{2}} = \alpha k \left( 1-\frac{1}{\mu \ell}\right)  = \alpha \frac{c_L}{6} \, , \qquad \sqrt{2 \bar{c}_{2}} = \bar{\alpha} k \left( 1+\frac{1}{\mu \ell}\right) =\bar{\alpha} \frac{c_R}{6} ~.
\ee
For the purpose of computing entanglement entropy (and hence mimicking the correct replica geometry at the boundary), the effect of the Wilson line should be to create a conical defect on the boundary metric. This is simply achieved by setting $\alpha = \bar{\alpha}$:  this sets $a_{L,\rm source} = a_{R, \rm source}$ and the corresponding conical metric reads
\be
{ds^2\over \ell^2} = d\rho^2 + e^{2\rho} \left( dr^2 + r^2 (\alpha -1)^2 d\theta^2 \right)~.
\ee
Here we have introduced polar coordinates $(r, \theta)$ on our Euclidean CFT spacetime by $z = r e^{i\theta}$ and $\bar{z} = r e^{-i \theta}.$  The strength of the conical singularity is governed by  $\alpha$ which should be identified with $n$ (or $\epsilon$) in the main text.  In the limit $n \rightarrow 1$ we find
\be\label{casrl}
\sqrt{2 c_{2}} = \frac{c_L}{6}(n-1) +O(n-1)^2\, , \quad \sqrt{2 \bar{c}_{2}} = \frac{c_R}{6} (n-1) +O(n-1)^2~.
\ee
This fixes the Casimirs. As expected, they reproduce the values of mass and spin in \eqref{eq:msmpd} upon using \eqref{eq:cms}.

 \vskip .2 in {\bf 2. Holographic entanglement entropy}  \vskip .2 in
Having constructed independent left and right Wilson lines, we can compute holographic entanglement entropy. Following the logic in \cite{Ammon:2013hba} we postulate that 
\bea\label{eedef}
S_{\rm EE} &=& \lim_{n\to 1}{1\over 1-n}\log\le(W_{\sR}(C)\ri) \cr
&=&\lim_{n\to 1}{1\over 1-n}\log\le(W_{L}(C)\ri) +\lim_{n\to 1}{1\over 1-n}\log\le(W_{R}(C)\ri)~,\label{eesl2}
\eea
when the endpoints of the curve $C$ are at the boundary of AdS$_3$. %
In the following we will explicitly evaluate \eqref{eesl2} in a semiclassical approximation.

 \vskip .2 in {\it a. Open path} \vskip .2 in

Consider first evaluating the action of the probe for an open interval in any locally AdS$_3$ background.\footnote{We remind the reader that we will be rather brief here. Details of the technique can be found in \cite{Ammon:2013hba}.} The path  $x^{\mu}(s)=(\rho(s), x^\pm(s))$ that characterizes $C$ satisfies the following boundary conditions
\be
\rho(s=s_f) = \rho(s=0)\equiv  \rho_0~,  \qquad x^\pm(s=s_f)-x^\pm(s=0)\equiv \Delta x^\pm~, \label{path}
\ee
 Here $s$ is the parameter along the path, varying from $s = 0$ to $s = s_f$. 
  
We start by solving for the dynamics of $U_L$ in \eqref{eq:actL}.  The solution with $A_L=0$ is
\be\label{eq:cc}
U_{L,0}(s) =  \exp\le( -\alpha(s) P_{L,0}\ri)u_0~,\quad
\ee
with $\dot P_{L,0}=0$, and the on-shell action is given by
\bea\label{eq:on}
S_{L,\rm on-shell} &=&- \int_C ds\, \dot \alpha(s) \Tr P^2_{L,0} =- c_2 (\alpha(s_f)-\alpha(0)) \equiv-c_2\Delta\alpha~. 
\eea

Since $A_L$ is a flat connection, to construct solutions with $A_L\neq 0$ we can gauge-up from $A_L=0$ to 
\be\label{eq:AL}
A_L= L dL^{-1}~,\quad  L(x^{\pm},\rho)= \exp\le(-\rho J_0\ri) \exp\le(-\int^{x}_{x_0} dx^i a_i \ri)~.
\ee
We have made a particular and conventional gauge choice for the $\rho$-dependence. Under \eqref{eq:AL}  our probe will transform as 
\be\label{eq:cd}
U_L(s) = L(s)U_{L,0}(s)~,
\ee
where $L(s)\equiv L(x^{\pm}(s),\rho(s))$, which is a solution to the equations of motion \eqref{eq:R1}. 
The appeal of this approach is that \eqref{eq:on} is still a valid expression when $A_L\neq0$, so the task at hand is to compute $\Delta\alpha$ for the set of backgrounds \eqref{eq:AL}. 

To do so we impose  boundary conditions on \eqref{eq:cd}; in particular, combining \eqref{eq:cd} with \eqref{eq:cc} we find
 \bea\label{eq:left}
 e^{-\Delta \alpha P_{L,0}} =  L(s_f)^{-1}U_{L}(s_f) U_{L}(0)^{-1}L(0) 
 \eea
 where $\Delta\alpha\equiv \alpha(s_f)-\alpha(0)$.  Equation \eqref{eq:left} determines $\Delta \alpha$ in terms of $A_L$ and the boundary data for $U_L$. More explicitly, taking the trace of \eqref{eq:left} gives
 \be\label{leftfinal}
 2 \cosh\le( \Delta\alpha \sqrt{ c_{2}\over 2}\ri) ={\rm Tr}_f\le(  L(0) L(s_f)^{-1} U_{L}(s_f)U_{L}(0)^{-1} \ri)~.
 \ee

 For the right invariant action the logic goes exactly the same with the appropriate replacements: provided a background connection of the form
 \be\label{eq:AR}
A_R= R^{-1}dR ~,\quad R(x^{\pm},\rho) = \exp\le( \int_{x_0}^{x} dx^i \bar a_i \ri) \exp\le(-\rho J_0\ri)~,
\ee
a solution to the equations of motion \eqref{eq:L1} is
\be
U_R(s)=U_{R,0} R(s)~, \quad U_{R,0}= u_0 \exp\le(- \bar \alpha(s) P_{R,0}\ri)~,\quad  {\rm Tr} P_{R,0}^2=\bar c_2 ~.
\ee
The right moving action \eqref{eq:actR} is simply
\be\label{eq:onR}
S_{R,\rm on-shell}=- \int_C ds\, \dot{\bar{\alpha}}(s) \Tr P^2_{R,0} =-\bar c_2 \Delta \bar \alpha~.
\ee
Analogously to \eqref{eq:left}--\eqref{leftfinal}, we solve for $\Delta\bar\alpha$ from the equation%
 \be\label{rightfinal}
 2 \cosh\le( \Delta\bar\alpha \sqrt{ \bar c_{2}\over 2}\ri) ={\rm Tr}_f\le( R(0)^{-1}R(s_f)U_R(s_f)^{-1} U_R(0)  \ri)~.
 \ee

To summarize, given a pair of flat connections in the gauge \eqref{eq:AL}, \eqref{eq:AR}, one computes the single interval entanglement entropy as follows. Equations \eqref{leftfinal} and \eqref{rightfinal} can  be solved for $\Delta\alpha$ and $\Delta\bar\alpha$. Combined with the Casimirs \eqref{casrl}, these define the on-shell actions via \eqref{eq:on} and \eqref{eq:onR}. In the semiclassical limit $W_{L,R}(C) \sim \exp\left(-S_{L,R,\rm on-shell}(C)\right)$, the on-shell actions in turn yield the entanglement entropy \eqref{eedef}. 

It may be verified that when applied to the backgrounds treated in the main text (Poincar\'e AdS and the rotating, planar BTZ black hole), with the appropriate choice of boundary conditions on $U_{L,R}$ at $s=s_f,0$, this prescription reproduces our earlier results.

 \vskip .2 in {\it b. Thermal entropy} \vskip .2 in

To end this appendix, we will compute $W_{L,R}(C)$ for a closed spatial curve; using \eqref{casrl} and \eqref{eesl2} the closed Wilson loop will reproduce the thermal entropy of the solution. Again, most of the derivations here follow from \cite{Ammon:2013hba}.

The main difference with respect to the open interval derivation is that $U(s)$ has to be a periodic function, $U_i=U_f$, and we take $\Delta \phi =2\pi$. Then using \eqref{eq:cc} and \eqref{eq:cd} we have
\bea
 \exp\le(\Delta \alpha P_{L,0}\ri) &= &L(s_f)^{-1} L(0) \cr
 &=& \exp(2\pi a_\phi) ~,
\eea
where we used that for a closed loop the combination $L(0)^{-1} L(s_f)$ is the holonomy around the spatial cycle $\phi\sim \phi +2\pi$.  Defining $\lambda_{L}$ and $\lambda_\phi$ as the diagonal matrices whose entries are the eigenvalues of $P_L$ and $a_\phi$ we find
\be
-\Delta \alpha \lambda_{L}= 2\pi \lambda_\phi~.
\ee
And using ${\rm tr}_f(\lambda_L J_0)=\sqrt{c_{2}\over 2}$  gives
\be
\Delta\alpha=-{2\pi\sqrt{2\over c_{2}} } {\rm tr}_f(\lambda_\phi J_0)
\ee
On the right, one proceeds analogously; the result is 
\be
\Delta\bar \alpha={2\pi \sqrt{2\over \bar c_{2}} } {\rm tr}_f(\bar\lambda_\phi J_0)~.
\ee
where $\bar\lambda_\phi$ is the diagonal matrix whose entries are the eigenvalues of $\bar a_\phi$.
The total action is the sum of \eqref{eq:on} and \eqref{eq:onR},
\bea\label{stot}
S_{\rm thermal}&=& S_{L, \rm on-shell}+S_{R, \rm on-shell}\cr
&=&2\pi  {c_L\over 6}  {\rm tr}_f(\lambda_\phi J_0)- 2\pi {c_R\over 6 } {\rm tr}_f(\bar\lambda_\phi J_0)
\eea
For $c_L=c_R$ this yields the formula for the thermal entropy in \cite{deBoer:2013gz}.  %

We can apply this formula for the BTZ black hole. The connection in this case is
\be\label{eq:btz} a = \le(J_1 - \frac{E_L}{k_L} J_{-1}\ri) dx^+ ~,\qquad \bar{a} = -\le(J_{-1} - \frac{E_R}{k_R}  J_1\ri)dx^-~,
\ee
where $k_{L,R}$ are defined in \eqref{tmgact}--\eqref{levels}. The identification of $E_L$ and $E_R$ with the mass $M$ and angular momentum $J$ of the black hole receives corrections from the Chern-Simons term \cite{Kraus:2005zm}. Using the relations \eqref{levels}, the charges can be written neatly as
\be\label{zeromodes}
E_L = {M-J\over 2} ~, \quad E_R = {M+J\over 2} 
\ee
where $(E_L,E_R)$ are the Virasoro zero mode energies. Upon combining \eqref{stot}--\eqref{zeromodes}, one recovers the correct BTZ thermal entropy \eqref{cardy} for arbitrary $(c_L,c_R)$.

\bibliographystyle{utphys}
\bibliography{all}

\end{document}